\documentclass[10pt,draftcls,onecolumn]{IEEEtran}
\renewcommand{\baselinestretch}{1.3}

\usepackage{times,latexsym,amsfonts,amsmath,amssymb,mathrsfs,verbatim,cite}
\usepackage{epsfig,epsf}
\usepackage{graphicx}
\usepackage[dvips]{color}
\usepackage{txfonts}

\def\zZ{{\mathbb Z}}
\def\rR{{\mathbb R}}

\def\pP{{\mathbb P}}

\def\QED{\mbox{\rule[0pt]{1.5ex}{1.5ex}}}

\def\@begintheorem#1#2{\tmpitemindent\itemindent\topsep 0pt\rm\trivlist
    \item[\hskip \labelsep{\indent\it #1\ #2:}]\itemindent\tmpitemindent}
\def\@opargbegintheorem#1#2#3{\tmpitemindent\itemindent\topsep 0pt\rm \trivlist
    \item[\hskip\labelsep{\indent\it #1\ #2\
    \rm(#3):}]\itemindent\tmpitemindent}
\def\@endtheorem{\endtrivlist\unskip}

\newtheorem{theorem}{Theorem}
\newtheorem{definition}{Definition}

\newtheorem{proposition}{Proposition}
\newtheorem{lemma}{Lemma}
\newtheorem{corollary}{Corollary}

\renewcommand{\theequation}{\arabic{section}.\arabic{equation}}

\newcommand{\supp}{\operatorname{supp}}

\begin{document}

\title{Distributed Source Coding for \\ Interactive Function Computation$^{\text{\small 1}}$}


\author{\authorblockN{Nan Ma and Prakash Ishwar}\\
\authorblockA{Department of Electrical and Computer Engineering \\
Boston University, Boston, MA 02215 \\ {\tt \{nanma, pi\}@bu.edu}}}

\maketitle

\begin{abstract}
A two-terminal interactive distributed source coding problem with
alternating messages for function computation at both locations is
studied.  For any number of messages, a computable characterization of
the rate region is provided in terms of single-letter information
measures. While interaction is useless in terms of the minimum
sum-rate for lossless source reproduction at one or both locations,
the gains can be arbitrarily large for function computation even when
the sources are independent. For a class of sources and functions,
interaction is shown to be useless, even with infinite messages, when
a function has to be computed at only one location, but is shown to be
useful, if functions have to be computed at both locations.  For
computing the Boolean AND function of two independent Bernoulli
sources at both locations, an achievable infinite-message sum-rate
with infinitesimal-rate messages is derived in terms of a
two-dimensional definite integral and a rate-allocation curve. A
general framework for multiterminal interactive function computation
based on an information exchange protocol which successively switches
among different distributed source coding configurations is developed.
%
%
For networks with a star topology, multiple rounds of interactive
coding is shown to decrease the scaling law of the total network rate
by an order of magnitude as the network grows.
%
%
\end{abstract}


\section{Introduction}
\addtocounter{footnote}{+1} \footnotetext{This material is based upon
work supported by the US National Science Foundation (NSF) under award
(CAREER) CCF--0546598.
Any opinions, findings, and conclusions or recommendations expressed
in this material are those of the authors and do not necessarily
reflect the views of the NSF. A part of this work was presented in ISIT'08.
}

In networked systems where distributed inferencing and control needs
to be performed, the raw-data (source samples) generated at different
nodes (information sources) needs to be transformed and combined in a
number of ways to extract actionable information.  This requires
performing distributed computations on the source samples. A pure
data-transfer solution approach would advocate first reliably
reproducing the source samples at decision-making nodes and then
performing suitable computations to extract actionable information.
Two-way interaction and statistical dependencies among source,
destination, and relay nodes, would be utilized, if at all, to
primarily improve the reliability of data-reproduction than overall
computation-efficiency.

However, to maximize the overall computation-efficiency, it is
necessary for nodes to interact bidirectionally, perform computations,
and exploit statistical dependencies in data as opposed to only
generating, receiving, and forwarding data. In this paper we attempt
to formalize this common wisdom through some examples of distributed
function-computation problems with the goal of minimizing the total
number of bits exchanged per source sample. Our objective is to
highlight the role of interaction in computation-efficiency within a
distributed source coding framework involving block-coding asymptotics
and vanishing probability of function-computation error.  We derive
information-theoretic characterizations of the set of feasible
coding-rates for these problems and explore the fascinating interplay
of function-structure, distribution-structure, and interaction.

\subsection{Problem setting} \label{subsec:introsetup}

\begin{figure}[!htb]
\begin{center}
\scalebox{0.5}{\input{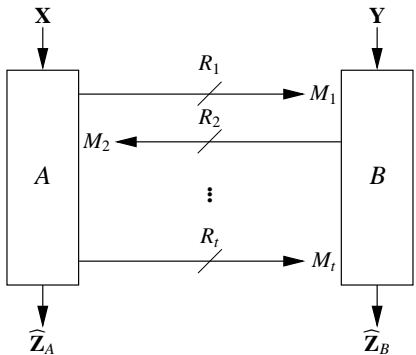}}
\end{center}
\caption{\sl Interactive distributed source coding with $t$ alternating
  messages.
\label{fig:structure}}
\end{figure}

Consider the following general two-terminal interactive distributed
source coding problem with alternating messages illustrated in
Figure~\ref{fig:structure}. Here, $n$ samples ${\mathbf X} := X^n :=
(X(1),\ldots,X(n)) \in {\mathcal X}^n$, of an information source are
available at location $A$. A different location $B$ has $n$ samples
${\mathbf Y} \in {\mathcal Y}^n$ of a second information source which
are statistically correlated to ${\mathbf X}$. Location $A$ desires to
produce a sequence $\widehat{{\mathbf Z}}_A \in {\mathcal Z}_A^n$ such
that $d_A^{(n)}({\mathbf X},{\mathbf Y},\widehat{{\mathbf Z}}_A) \leq
D_A$ where $d_A^{(n)}$ is a nonnegative distortion function of $3n$
variables. Similarly, location $B$ desires to produce a sequence
$\widehat{{\mathbf Z}}_B \in {\mathcal Z}_B^n$ such that
$d_B^{(n)}({\mathbf X},{\mathbf Y},\widehat{{\mathbf Z}}_B) \leq
D_B$. All alphabets are assumed to be finite.  To achieve the desired
objective, $t$ coded messages, $M_1,\ldots,M_t$, of respective bit
rates (bits per source sample), $R_1,\ldots,R_t$, are sent alternately
from the two locations starting with location $A$ of location $B$.
The message sent from a location can depend on the source samples at
that location and on all the previous messages (which are available to
both locations). There is enough memory at both locations to store all
the source samples and messages.  An important goal is to characterize
the set of all rate $t$-tuples ${\mathbf R}:= (R_1,\ldots,R_t)$ for
which both $\pP(d_A^{(n)}({\mathbf X},{\mathbf Y},\widehat{{\mathbf
    Z}}_A) > D_A)$ and $\pP(d_B^{(n)}({\mathbf X},{\mathbf
  Y},\widehat{{\mathbf Z}}_B) > D_B)$ $\rightarrow 0$ as $n
\rightarrow \infty$. This set of rate-tuples is called the rate
region.

\subsection{Related work}

The available literature closely related to this problem can be
roughly partitioned into three broad categories. The salient features
of related problems in these three categories are summarized below
using the notation of the problem setting described above.

\subsubsection{Communication complexity~\cite{CommComplexity}}

Here, ${\bf X}$ and ${\bf Y}$ are typically deterministic, $t$ is not
fixed in advance, and $d_A^{(n)}$ and $d_B^{(n)}$ are the indicator
functions of the sets $\{\widehat{{\mathbf Z}}_A \neq {\mathbf
  f}_A({\mathbf X},{\mathbf Y})\}$ and $\{\widehat{{\mathbf Z}}_B \neq
{\mathbf f}_B({\mathbf X},{\mathbf Y})\}$ respectively.  Thus, the
goal is to compute the function ${\mathbf f}_A({\mathbf X},{\mathbf
  Y})$ at location $A$ and the function ${\mathbf f}_B({\mathbf
  X},{\mathbf Y})$ at location $B$. Both deterministic and randomized
coding strategies have been studied.  If coding is deterministic, the
functions are required to be computed {\em without error}, i.e., $D_A
= D_B = 0$. If coding is randomized, with the sources of randomness
independent of each other and ${\mathbf X}$ and ${\mathbf Y}$, then
$\widehat{{\mathbf Z}}_A$ and $\widehat{{\mathbf Z}}_B$ are random
variables. In this case, computation could be required to be
error-free and the termination time $t$ random (the Las-Vegas
framework) or the termination time $t$ could be held fixed but large
enough to keep the probability of computation error smaller than some
desired value (the Monte-Carlo framework).

The coding-efficiency for function computation is called communication
complexity. When coding is deterministic, communication complexity is
measured in terms of the minimum value, over all codes, of the total
number of bits that need to be exchanged between the two locations, to
compute the functions without error, irrespective of the values of the
sources. When coding is randomized, both the worst-case and the
expected value of the total number of bits, over all sources of
randomization, have been considered. The focus of much of the
literature has been on establishing order-of-magnitude upper and lower
bounds for the communication complexity and not on characterizing the
set of all source coding rate tuples in bits per source sample. In
fact, the ranges of ${\mathbf f}_A$ and ${\mathbf f}_B$ considered in
the communication complexity literature are often orders of magnitude
smaller than their domains.  This would correspond to a vanishing
source coding rate.

Recently, however, Giridhar and Kumar successfully applied the
communication complexity framework to study how the rate of function
computation can scale with the size of the network for deterministic
sources \cite{Kumar2005,Kumar2006}. They considered a network where
each node observes a (deterministic) sequence of source samples and a
sink node where the sequence of function values needs to be
computed. To study how the computation rate scales with the network
size, they considered the class of connected random planar networks
and the class of co-located networks and focused on the divisible and
symmetric families of functions.

\subsubsection{Interactive source reproduction}

Kaspi~\cite{Kaspi1985} considered a distributed block source
coding~\cite[Section~14.9]{CoverThomas} formulation of this problem
for discrete memoryless stationary sources taking values in finite
alphabets. However, the focus was on source reproduction with
distortion and not function computation. The source reproduction
quality was measured in terms of two single-letter distortion
functions of the form $d_A^{(n)}({\mathbf x},{\mathbf y},\hat{{\mathbf
    z}}_A) := (1/n)\sum_{i=1}^nd_A(y(i), \hat{z}_{A}(i))$ and
$d_B^{(n)}({\mathbf x},{\mathbf y},\hat{{\mathbf z}}_B) :=
(1/n)\sum_{i=1}^nd_B(x(i), \hat{z}_{B}(i))$.  Coupled single-letter
distortion functions of the form $d_A(x(i), y(i), \hat{z}_A(i))$ and
$d_B(x(i), y(i), \hat{z}_B(i))$, and probability of block error for
lossless reproduction, were not considered.  For a fixed number of
messages $t$, a single-letter characterization of the {\em sum-rate
  pair} $(\sum_{j \ \text{odd}} R_j,\sum_{j \ \text{even}} R_j)$ (not
the entire rate region) was derived.  However, no examples were
presented to illustrate the benefits of two-way source coding. The key
question: ``does two-way (interactive) distributed source coding with
more messages require a strictly less sum-rate than with fewer
messages?''  was left unanswered.

The recent paper by Yang and He \cite{Dakehe} studied two-terminal
interactive source coding for the {\em lossless} reproduction of a
stationary non-ergodic source ${\bf X}$ at $B$ with decoder
side-information ${\bf Y}$.  Here, the code termination criterion
depended on the sources and previous messages so that $t$ was a random
variable.  Two-way interactive coding was shown to be strictly better
than one-way non-interactive coding.

\subsubsection{Interactive function computation}

In~\cite{Yamamoto1982}, Yamamoto studied the problem where $({\bf X},
{\bf Y})$ is a doubly symmetric binary source,\footnote{$(X(i),Y(i))
  \sim$ iid $p_{XY}(x,y) = 0.5(1-p)\delta_{xy} + 0.5p(1-\delta_{xy})$,
  where $\delta_{ij}$ is the Kronecker delta, and $x,y \in
  \{0,1\}$. We say $(X,Y) \sim$ DSBS$(p)$.} terminal $B$ is required
to compute a Boolean function of the sources satisfying an {\em
  expected} per-sample Hamming distortion criterion corresponding to
$d_B^{(n)}({\mathbf x},{\mathbf y},\hat{{\mathbf z}}_B) :=
(1/n)\sum_{i=1}^n (f_B(x(i),y(i))\oplus\hat{z}_{B}(i))$, where
$f_B(x,y)$ is a Boolean function, only one message is allowed, i.e.,
$t = 1$, and nothing is required to be computed at terminal $A$, i.e.,
$d_A^{(n)} = 0$.  This is equivalent to Wyner-Ziv source
coding~\cite{CoverThomas} with decoder side-information for a
per-sample distortion function which depends on the decoder
reconstruction and both the sources.  Yamamoto computed the
rate-distortion function for all the $16$ Boolean functions of two
binary variables and showed that they are of only three forms.

In~\cite{Han}, Han and Kobayashi studied a three-terminal problem
where ${\bf X}$ and ${\bf Y}$ are discrete memoryless stationary
sources taking values in finite alphabets, ${\bf X}$ is observed at
terminal one and ${\bf Y}$ at terminal two and terminal three wishes
to compute a samplewise function of the sources losslessly. Only
terminals one and two can each send only one message to terminal
three. Han and Kobayashi characterized the class of functions for
which the rate region of this problem coincides with the
Slepian-Wolf~\cite{CoverThomas} rate region.

Orlitsky and Roche~\cite{OrlitskyRoche} studied a distributed block
source coding problem whose setup coincides with Kaspi's
problem~\cite{Kaspi1985} described above. However, the focus was on
computing a samplewise function ${\bf f}_B({\bf X},{\bf Y}) =
(f_B(X(i),Y(i)))_{i=1}^n$ of the two sources at terminal
$B$ using up to two messages ($t \leq 2$). Nothing was required to be
computed at terminal $A$, i.e., $d_A^{(n)} = 0$. Both probability of
block error $\pP(\{\widehat{{\mathbf Z}}_B \neq {\mathbf f}_B({\mathbf
  X},{\mathbf Y})\})$ and per-sample expected Hamming distortion
$(1/n)\sum_{i=1}^n \pP(\hat{Z}_{B}(i) \neq f_B(X(i),Y(i)))$ were
considered.  A single-letter characterization of the rate region was
derived. Example~8 in \cite{OrlitskyRoche} showed that the sum-rate
with $2$ messages is strictly smaller than with one message.

\subsection{Contributions}

We study the two-terminal interactive function computation problem
described in Section~\ref{subsec:introsetup} for discrete memoryless
stationary sources taking values in finite alphabets. The goal is to
compute samplewise functions at one or both locations and the two
functions can be the same or different. We focus on a distributed
block source coding formulation involving a probability of block error
which is required to vanish as the blocklength tends to infinity. We
derive a computable characterization of the the rate region and the
minimum sum-rate for {\em any finite number of messages} in terms of
single-letter information quantities (Theorem~\ref{thm:rateregion} and
Corollary~\ref{cor:lowerbound}).  We show how the rate-regions for
different number of messages and different starting locations are
nested (Proposition~\ref{prop:struct}).  We show how the Markov chain
and conditional entropy constraints associated with the rate region
are related to certain geometrical properties of the support-set of
the joint distribution and the function-structure
(Lemma~\ref{lem:rectangle}). This relationship provides a link to the
concept of monochromatic rectangles which has been studied in the
communication complexity literature. We also consider a concurrent
kind of interaction where messages are exchanged simultaneously and
show how the minimum sum-rate is bounded by the sum-rate for
alternating-message interaction (Proposition~\ref{prop:concurrent}).
We also consider per-sample average distortion criteria based on {\em
  coupled} single-letter distortion functions which involve the
decoder output and both sources. For expected distortion as well as
probability of excess distortion we discuss how the single-letter
characterization of the rate-distortion region is related to the rate
region for probability of block error (Section~\ref{sec:RD}).

Striking examples are presented to show how the benefit of interactive
coding depends on the function-structure, computation at one/both
locations, and the structure of the source distribution. Interactive
coding is useless (in terms of the minimum sum-rate) if the goal is
{\em lossless source reproduction} at one or both locations but the
gains can be arbitrarily large for computing nontrivial functions
involving both sources even when the sources are independent
(Sections~\ref{subsec:sourceoneloc}, \ref{subsec:sourcetwoloc}, and
\ref{subsec:functiononeloc}).  For certain classes of sources and
functions, interactive coding is shown to have no advantage
(Theorems~\ref{thm:onereconnohelp} and \ref{thm:nohelpgeneral}).  In
fact, for doubly symmetric binary sources, interactive coding, with
even an unbounded number of messages is useless for computing {\em any
  function} at one location (Section~\ref{subsec:DSBSoneloc}) but {\em
  is useful} if computation is desired at both locations
(Section~\ref{subsec:DSBStwoloc}). For independent Bernoulli sources,
when the Boolean AND function is required to be computed at both
locations, we develop an achievable {\em infinite-message} sum-rate
with an infinitesimal rate for each message
(Section~\ref{subsec:integralrate}). This sum-rate is expressed in
analytic closed-form, in terms of two two-dimensional definite
integrals, which represent the total rate flowing in each direction,
and a rate-allocation curve which coordinates the progression of
function computation.

We develop a general formulation of multiterminal interactive function
computation in terms an interaction protocol which switches among many
distributed source coding configurations
(Section~\ref{sec:multiterminal}). We show how results for the
two-terminal problem can be used to develop insights into optimum
topologies for information flow in larger networks through a linear
program involving cut-set lower bounds
(Sections~\ref{subsec:cutsetbnd} and \ref{subsec:multiexamples}).  We
show that allowing any arbitrary number of interactive message
exchanges over multiple rounds cannot reduce the minimum total rate
for the K\"{o}rner-Marton problem~\cite{KornerMarton}. For networks
with a star topology, however, we show that interaction can, in fact,
decrease the scaling law of the total network rate by an order of
magnitude as the network grows (Example~3 in
Section~\ref{subsec:multiexamples}).

{\em Notation:} In this paper, the terms terminal, node, and location,
are synonymous and are used interchangeably.  The acronym `iid' stands
for independent and identically distributed and `pmf' stands for
probability mass function.  Boldface letters such as, ${\mathbf
  x},~{\mathbf X}$, etc., are used to denote vectors.  Although the
dimension of a vector is suppressed in this notation, it will be
clear from the context. With the exception of the symbols $R, D, N,
L, A$, and $B$, random quantities are denoted in upper case, e.g.,
$X,~{\mathbf X}$, etc., and their specific instantiations are
denoted in lower case, e.g., $X = x,~{\mathbf X} = {\mathbf x}$,
etc. When $X$ denotes a random variable, $X^n$ denotes the ordered
tuple $(X_1,\ldots,X_n)$ and $X_m^n$ denotes the ordered tuple
$(X_m,\ldots,X_n)$. However, for a set ${\mathcal S}$, ${\mathcal
S}^n$ denotes the $n$-fold Cartesian product ${\mathcal S} \times
\ldots \times {\mathcal S}$. The symbol $X(i-)$ denotes
$(X(1),\ldots,X(i-1))$ and $X(i+)$ denotes $(X(i+1),\ldots,X(n))$.
The indicator function of set $\mathcal{S}$ which is equal to one if
$x\in \mathcal{S}$ and is zero otherwise, is denoted by
$1_\mathcal{S}(x)$. The support-set of a pmf $p$ is the set over
which it is strictly positive and is denoted by $\supp(p)$. Symbols
$\oplus, \wedge$, and $\vee$ represent Boolean XOR, AND, and OR
respectively.


\section{\label{sec:problem}Two-terminal interactive function computation}

\subsection{\label{subsec:code}Interactive distributed source code}

We consider two statistically dependent discrete memoryless stationary
sources taking values in finite alphabets. For $i=1,\ldots,n$, let
$(X(i),Y(i))\sim$ iid $p_{XY}(x,y), x\in \mathcal{X}, y\in
\mathcal{Y}, |\mathcal{X}|<\infty, |\mathcal{Y}|<\infty$.  Here,
$p_{XY}$ is a joint pmf which describes the statistical dependencies
among the samples observed at the two locations at each time instant
$i$. Let $f_A: {\mathcal X} \times {\mathcal Y} \rightarrow {\mathcal
Z}_A$ and $f_B: {\mathcal X} \times {\mathcal Y} \rightarrow {\mathcal
Z}_B$ be functions of interest at locations $A$ and $B$ respectively,
where ${\mathcal Z}_A$ and ${\mathcal Z}_B$ are finite alphabets. The
desired outputs at locations $A$ and $B$ are ${\mathbf Z}_A$ and
${\mathbf Z}_B$ respectively, where for $i=1,\ldots,n$, $Z_A(i) :=
f_A(X(i), Y(i))$ and $Z_B(i) := f_B(X(i), Y(i))$.

\begin{definition}\label{def:code}
A (two-terminal) interactive distributed source code (for function
computation) with initial location $A$ and parameters
$(t,n,|{\mathcal M}_1|,\ldots,|{\mathcal M}_t|)$ is the tuple
$(e_1,\ldots,e_t,g_A,g_B)$ of $t$ block encoding functions
$e_1,\ldots,e_t$ and two block decoding functions $g_A, g_B$, of
blocklength $n$, where for $j=1,\ldots,t$,
\begin{eqnarray*}
(\mbox{Enc.}j) &e_j:& \left\{
\begin{array}{r@{\ , \quad}l}
{\mathcal X}^{n} \times \bigotimes_{i=1}^{j-1} {\mathcal M}_i
\rightarrow {\mathcal M}_j &\mbox{if }j\mbox{ is odd}\\ {\mathcal
Y}^{n} \times \bigotimes_{i=1}^{j-1} {\mathcal M}_i \rightarrow
{\mathcal M}_j &\mbox{if }j\mbox{ is even}
\end{array}
\right.,\\
 (\mbox{Dec.}A) &g_A:&{\mathcal X}^{n} \times
\bigotimes_{j=1}^{t} {\mathcal M}_j \rightarrow {\mathcal Z}_A^n,\\
(\mbox{Dec.}B) &g_B:&{\mathcal Y}^{n} \times \bigotimes_{j=1}^{t}
{\mathcal M}_j \rightarrow {\mathcal Z}_B^n.
\end{eqnarray*}
The output of $e_j$, denoted by $M_j$, is called the $j$-th message,
and $t$ is the number of messages. The outputs of $g_A$ and $g_B$
are denoted by $\widehat{\mathbf Z}_A$ and $\widehat{\mathbf Z}_B$
respectively. For each $j$, $(1/n) \log_2 |{\mathcal M}_j|$ is
called the $j$-th block-coding rate (in bits per sample).
\end{definition}

Intuitively speaking, $t$ coded messages, $M_1,\ldots,M_t$, are sent
alternately from the two locations starting with location $A$. The
message sent from a location can depend on the source samples at
that location and on all the previous messages (which are available
to both locations from previous message transfers). There is enough
memory at both locations to store all the source samples and
messages.

We consider two types of fidelity criteria for interactive function
computation in this paper. These are 1) probability of block error and
2) per-sample distortion.

\subsection{\label{subsec:proberror}Probability of block error and operational rate region}

Of interest here are the probabilities of block error $\pP(\mathbf
Z_A \neq \mathbf {\widehat Z}_A)$ and $\pP(\mathbf Z_B \neq \mathbf
{\widehat Z}_B)$ which are multi-letter distortion functions. The
performance of $t$-message interactive coding for function
computation is measured as follows.

\begin{definition}\label{def:rateregion}
A rate tuple ${\mathbf R} = (R_1, \ldots, R_t)$ is admissible for
$t$-message interactive function computation with initial location
$A$ if, $\forall \epsilon > 0$, $\exists~ N(\epsilon,t)$ such that
$\forall n> N(\epsilon,t)$, there exists an interactive distributed
source code with initial location $A$ and parameters
$(t,n,|{\mathcal M}_1|,\ldots,|{\mathcal M}_t|)$ satisfying
\begin{eqnarray*}
&&\frac{1}{n}\log_2 |{\mathcal M}_j| \leq R_j + \epsilon,\ j =
1,\ldots, t,\\
&& \pP({\mathbf Z}_A \neq \widehat{\mathbf Z}_A) \leq
\epsilon,\ \pP({\mathbf Z}_B \neq \widehat{\mathbf Z}_B) \leq \epsilon.
\end{eqnarray*}
\end{definition}

The set of all admissible rate tuples, denoted by ${\mathcal
R}^A_t$, is called the operational rate region for $t$-message
interactive function computation with initial location $A$. The rate
region is closed and convex due to the way it has been defined. The
minimum sum-rate $R^A_{sum,t}$ is given by $\min \left(\sum_{j=1}^t
R_j\right)$ where the minimization is over ${\mathbf R} \in
{\mathcal R}^A_t$. For initial location $B$, the rate region and the
minimum sum-rate are denoted by ${\mathcal R}^B_t$ and $R^B_{sum,t}$
respectively.

\subsection{\label{subsec:distortion}Per-sample distortion and operational rate-distortion region}

Let $d_A: \mathcal{X}\times\mathcal {Y} \times \mathcal
{Z}_A\rightarrow \rR^+$ and $d_B: \mathcal{X}\times\mathcal {Y} \times
\mathcal {Z}_B\rightarrow \rR^+$ be bounded single-letter distortion
functions. The fidelity of function computation can be measured by the
per-sample average distortion
\[d_A^{(n)}(\mathbf x, \mathbf y, \hat{\mathbf z}_A):=\frac{1}{n}\sum_{i=1}^{n}
d_A(x(i), y(i), \hat{z}_A(i)).\]
\[d_B^{(n)}(\mathbf x, \mathbf y, \hat{\mathbf z}_B):=\frac{1}{n}\sum_{i=1}^{n}
d_B(x(i), y(i), \hat{z}_B(i)).\] Of interest here are either the
expected per-sample distortions $E[d_A^{(n)}(\mathbf X, \mathbf Y,
  \widehat{\mathbf Z}_A)]$ and $E[d_B^{(n)}(\mathbf X, \mathbf Y,
  \widehat{\mathbf Z}_B)]$ or the probabilities of excess distortion
$\pP(d_A^{(n)}(\mathbf X, \mathbf Y, \widehat{\mathbf Z}_A) > D_A)$
and $\pP(d_B^{(n)}(\mathbf X, \mathbf Y, \widehat{\mathbf Z}_B) >
D_B)$. Note that although the desired functions $f_A$ and $f_B$ do not
explicitly appear in these fidelity criteria, they are subsumed by
$d_A$ and $d_B$ because they accommodate general relationships between
the sources and the outputs of the decoding functions. The performance
of $t$-message interactive coding for function computation is measured
as follows.

\begin{definition}\label{def:ratedistortion}
A rate-distortion tuple ${(\mathbf R,\mathbf D)} = (R_1, \ldots,
R_t, D_A, D_B)$ is admissible for $t$-message interactive function
computation with initial location $A$ if, $\forall \epsilon > 0$,
$\exists~ N(\epsilon,t)$ such that $\forall n> N(\epsilon,t)$, there
exists an interactive distributed source code with initial location
$A$ and parameters $(t,n,|{\mathcal M}_1|,\ldots,|{\mathcal M}_t|)$
satisfying
\begin{eqnarray*}
&&\frac{1}{n}\log_2 |{\mathcal M}_j| \leq R_j + \epsilon,\ j =
1,\ldots, t,\\
&& E[d_A^{(n)}(\mathbf X, \mathbf Y, \widehat{\mathbf Z}_A)]\leq D_A
+ \epsilon,\ E[d_B^{(n)}(\mathbf X, \mathbf Y, \widehat{\mathbf
Z}_B)] \leq D_B+\epsilon.
\end{eqnarray*}
\end{definition}

The set of all admissible rate-distortion tuples, denoted by
${\mathcal {RD}}^A_t$, is called the operational rate-distortion
region for $t$-message interactive function computation with initial
location $A$. The rate-distortion region is closed and convex due to
the way it has been defined. The sum-rate-distortion function
$R^A_{sum,t}(\mathbf D)$ is given by $\min \left(\sum_{j=1}^t
R_j\right)$ where the minimization is over all $\mathbf R$ such that
$(\mathbf R, \mathbf D) \in \mathcal {RD}^A_t$. For initial location
$B$, the rate-distortion region and the minimum sum-rate-distortion
function are denoted by ${\mathcal {RD}}^B_t$ and
$R^B_{sum,t}(\mathbf{D})$ respectively.

The admissibility of a rate-distortion tuple can also be defined in
terms of the probability of excess distortion by replacing the
expected distortion conditions in Definition~\ref{def:ratedistortion}
by the conditions $\pP(d_A^{(n)}(\mathbf X, \mathbf Y,
\widehat{\mathbf Z}_A) > D_A) \leq \epsilon$ and
$\pP(d_B^{(n)}(\mathbf X, \mathbf Y, \widehat{\mathbf Z}_B) > D_B)
\leq \epsilon$. Although these conditions appear to be more
stringent\footnote{Any tuple which is admissible according to the
probability of excess distortion criteria is also admissible according
to the expected distortion criteria.}, it can be shown\footnote{Using
strong-typicality arguments in the proof of the achievability part of
the single-letter characterization of the rate-distortion region.}
that they lead to the same operational rate-distortion region.  For
simplicity, we focus on the expected distortion conditions as in
Definition~\ref{def:ratedistortion}.

\subsection{\label{subsec:definitioncomments}Discussion}

For a $t$-message interactive distributed source code, if $|{\mathcal
M}_t|=1$, then $M_t = constant$ (null message) and nothing needs to be
sent in the last step and the $t$-message code reduces to a
$(t-1)$-message code.  Thus the $(t-1)$-message rate region is
contained within the $t$-message rate region.  For generality and
convenience, $|{\mathcal M}_j|=1$ is allowed for all $j \leq t$. The
following proposition summarizes some key properties of the rate
regions which are needed in the sequel.

\begin{proposition}\label{prop:struct} (i) If $(R_1,\ldots,R_{t-1}) \in \mathcal R^A_{t-1}$, then
$(R_1,\ldots,R_{t-1},0) \in \mathcal R^A_{t}$. Hence $R^A_{sum,(t-1)}
\geq R^A_{sum,t}$. (ii) If $(R_1,\ldots,R_{t-1}) \in \mathcal
R^B_{t-1}$, then $(0,R_1,\ldots,R_{t-1}) \in \mathcal R^A_{t}$. Hence
$R^B_{sum,(t-1)} \geq R^A_{sum,t}$. Similarly, $R^A_{sum,(t-1)} \geq
R^B_{sum,t}$. (iii) $\lim_{t \rightarrow \infty} R^A_{sum,t} = \lim_{t
\rightarrow \infty} R^B_{sum,t} =: R_{sum,\infty}$.
\end{proposition}

{\em Proof:} (i) Any $(t-1)$-message code with initial location $A$
can be regarded as a special case of a $t$-message code with initial
location $A$ by taking $|{\mathcal M}_{t-1}|=1$. (ii) Any
$(t-1)$-message code with initial location $B$ can be regarded as a
special case of a $t$-message code with initial location $A$ by taking
$|{\mathcal M}_{1}|=1$. (iii) From (i), $R^A_{sum,t}$ and
$R^B_{sum,t}$ are nonincreasing in $t$ and bounded from below from
zero, so the limits exist. From (ii), $R^A_{sum,(t-1)} \geq
R^B_{sum,t} \geq R^A_{sum,(t+1)}$, hence the limits are equal.
\hspace*{\fill}~\QED

Proposition~\ref{prop:struct} is also true for any fixed distortion
levels $(D_A, D_B)$ if we replace rate regions and minimum sum-rates
in the proposition by rate-distortion regions and
sum-rate-distortion functions respectively.


\subsection{\label{subsec:concurrent}Interaction with concurrent message exchanges}

In contrast to the type of interaction described in
Section~\ref{subsec:code} which involves \emph{alternating message
  transfers}, one could also consider another type of interaction
which involves \emph{concurrent messages exchanges}. In this type of
interaction, in the $j$-th round of interaction, two messages
$M^{AB}_j$ and $M^{BA}_j$ are generated simultaneously by encoding
functions $e^{AB}_j$ (at location $A$) and $e^{BA}_j$ (at location
$B$) respectively. These messages are based on the source samples
which are available at each location and on all the previous messages
$\{ M^{AB}_i,M^{BA}_i\}_{i=1}^{j-1}$ which are available to both
locations from previous rounds of interaction. Then $M^{AB}_j$ and
$M^{BA}_j$ are exchanged. In $t$ rounds, $2t$ messages are
transferred. After $t$ rounds of interaction, decoding functions $g_A$
and $g_B$ generate function estimates based on all the messages and
the source samples which are available at locations $A$ and $B$
respectively. We can define the rate region and the rate-distortion
region for concurrent interaction as in
Sections~\ref{subsec:proberror} and \ref{subsec:distortion} for
alternating interaction. Let $R_{sum,t}^{conc}$ denote the minimum
sum-rate for $t$-round interactive function computation with
concurrent message exchanges.

The following proposition shows how the minimum sum-rates for
concurrent and alternating types of interaction bound each other.
This is based on a purely structural comparison of alternating and
concurrent modes of interaction.

\begin{proposition}\label{prop:concurrent}
(i) $R^A_{sum,t}\geq R_{sum,t}^{conc}\geq R^A_{sum,(t+1)}$. (ii)
$\lim_{t\rightarrow \infty} R_{sum,t}^{conc}= \lim_{t\rightarrow
\infty} R_{sum,t}^{A}=R_{sum,\infty}$.
\end{proposition}

{\em Proof:} (i) The first inequality holds because any $t$-message
interactive code with alternating messages and initial location $A$
can be regarded as a special case of a $t$-round interactive code
with concurrent messages by taking $|\mathcal M_j^{AB}|=1$ for all
even $j$ and $|\mathcal M_j^{BA}|=1$ for all odd $j$.

The second inequality can be proved as follows. Given any $t$-round
interactive code with concurrent messages and encoding functions
$\{e^{AB}_j,e^{BA}_j\}_{j=1}^{t}$, one can construct a
$(t+1)$-message interactive code with alternating messages as
follows: (1) Set $e_1:=e^{AB}_1$. (2) For $j=2,\ldots,t$, if $j$ is
even, define $e_j$ as the combination of $e^{BA}_{j-1}$ and
$e^{BA}_{j}$, otherwise, define $e_j$ as the combination of
$e^{AB}_{j-1}$ and $e^{AB}_{j}$. (3) If $t$ is even, set
$e_{t+1}:=e^{AB}_t$, otherwise set $e_{t+1}:=e^{BA}_t$. It can be
verified by induction that the inputs of $\{e_1,\ldots,e_{t+1}\}$
defined in this way are indeed available when these encoding
functions are used. Hence these are valid encoding functions for
interactive coding with alternating messages. This $(t+1)$-message
interactive code with alternating messages has the same sum-rate as
the original $t$-round interactive code with concurrent messages.
Therefore we have $R^A_{sum,(t+1)}\leq R_{sum,t}^{conc}$.

(ii) This follows from (i). \hspace*{\fill}~\QED

Although a $t$-round interactive code with concurrent messages uses
$2t$ messages, the sum-rate performance is bounded by that of an
alternating-message code with only $(t+1)$ messages. When $t$ is
large, the benefit of concurrent interaction over alternating
interaction disappears. Due to this reason and because for
two-terminal function computation it is easier to describe results for
alternating interaction, in Sections~\ref{sec:rateregions} and
\ref{sec:examples} our discussion will be confined to alternating
interaction. For multiterminal function computation, however, the
framework of concurrent interaction becomes more convenient. Hence in
Section~\ref{sec:multiterminal} we consider multiterminal function
computation problems with concurrent interaction.
%

\section{Rate region}\label{sec:rateregions}

\subsection{Probability of block error}

When the probability of block error is used to measure the quality of
function computation, the rate region for $t$-message interactive
distributed source coding with alternating messages can be
characterized in terms of single-letter mutual information quantities
involving auxiliary random variables satisfying conditional entropy
constraints and Markov chain constraints. This characterization is
provided by Theorem~\ref{thm:rateregion}.

\begin{theorem}\label{thm:rateregion}
\begin{eqnarray}
\lefteqn{{\mathcal R}_t^A = \{~\mathbf R~|~\exists \ U^t,
s.t.\  \forall i=1,\ldots,t,}&&\nonumber\\
&&R_i \geq \left\{
\begin{array}{cc}
I(X;U_i|Y, U^{i-1}),~U_i - (X, U^{i-1})- Y, & i\mbox{ odd} \\
I(Y;U_i|X, U^{i-1}),~U_i - (Y, U^{i-1})- X, & i\mbox{ even}
\end{array}
\right. \nonumber \\
&& H(f_A(X,Y)|X,U^t)=0, \
H(f_B(X,Y)|Y,U^t)=0~\},\label{eqn:rateregion}
\end{eqnarray}
where $U^t$ are auxiliary random variables taking values in
alphabets with the cardinalities bounded as follows,
\begin{equation}
|\mathcal{U}_j| \leq \left\{
\begin{array}{cc}
|\mathcal{X}| \left(\prod_{i=1}^{j-1} |\mathcal{U}_i|\right) + t -j + 3 , & j\mbox{ odd}, \\
|\mathcal{Y}| \left(\prod_{i=1}^{j-1} |\mathcal{U}_i|\right) + t - j
+ 3, & j\mbox{ even}.
\end{array}
\right.\label{eqn:cardinality}
\end{equation}
\end{theorem}

It should be noted that the right side of (\ref{eqn:rateregion}) is
convex and closed. This is because $\mathcal{R}^A_t$ is convex and
closed and Theorem~\ref{thm:rateregion} shows that the right side of
(\ref{eqn:rateregion}) is the same as $\mathcal{R}^A_t$. In fact the
convexity and closedness of the right side of (\ref{eqn:rateregion})
can be shown directly without appealing to
Theorem~\ref{thm:rateregion} and the properties of $\mathcal{R}^A_t$.
This is explained at the end of Appendix~\ref{app:converse}.

The proof of achievability follows from standard random coding and
random binning arguments as in the source coding with side information
problem studied by Wyner, Ziv, Gray, Ahlswede, and
K\"{o}rner~\cite{CoverThomas} (also see Kaspi \cite{Kaspi1985}). We
only develop the intuition and informally sketch the steps leading to
the proof of achievability. The key idea is to use a sequence of
``Wyner-Ziv-like'' codes. First, Enc.1 quantizes $\mathbf X$ to
$\mathbf U_1 \in (\mathcal{U}_1)^n$ using a random codebook-1. The
codewords are further randomly distributed into bins and the bin index
of $\mathbf{U}_1$ is sent to location $B$. Enc.2 identifies
$\mathbf{U}_1$ from the bin with the help of $\mathbf Y$ as decoder
side-information. Next, Enc.2 jointly quantizes $(\mathbf Y, \mathbf
U_1)$ to $\mathbf U_2 \in (\mathcal{U}_2)^n$ using a random
codebook-2. The codewords are randomly binned and the bin index of
$\mathbf U_2$ is sent to location $A$. Enc.3 identifies $\mathbf U_2$
from the bin with the help of $(\mathbf X, \mathbf U_1)$ as decoder
side-information. Generally, for the $j$-th message, $j$ odd, Enc.$j$
jointly quantizes $(\mathbf X, \mathbf U^{j-1})$ to $\mathbf U_j \in
(\mathcal{U}_j)^n$ using a random codebook-$j$. The codewords are
randomly binned and the bin index of $\mathbf U_j$ is sent to location
$B$. Enc.$(j+1)$ identifies $\mathbf U_j$ from the bin with the help
of $(\mathbf Y, \mathbf U^{j-1})$ as decoder side information. If $j$
is even, interchange the roles of locations $A$ and $B$ and sources
$\mathbf X$ and $\mathbf Y$ in the procedure for an odd $j$. Note that
$H(f_A(X,Y)|X,U^t)=0$ implies the existence of a deterministic
function $\phi_A$ such that $\phi_A(X,U^t)=f_A(X,Y)$. At the end of
$t$ messages, Dec.$A$ produces $\mathbf {\widehat Z}_A$ by $\widehat
Z_A(i)=\phi_A(X(i),U^t(i)), \forall i=1,\ldots,n$. Similarly, Dec.$B$
produces $\mathbf {\widehat Z}_B$. The rate and Markov chain
constraints ensure that all quantized codewords are jointly strongly
typical with the sources and are recovered with a probability which
tends to one as $n \rightarrow \infty$. The conditional entropy
constraints ensure that the corresponding block error probabilities
for function computation go to zero as the blocklength tends to
infinity.

%

The (weak) converse is proved in Appendix \ref{app:converse}
following \cite{Kaspi1985} using standard information inequalities,
suitably defining auxiliary random variables, and using
convexification (time-sharing) arguments. The conditional entropy
constraints are established using Fano's inequality as in
\cite[Lemma~1]{Han}. The proof of cardinality bounds for the
alphabets of the auxiliary random variables is also sketched.
\hspace*{\fill}~\QED

\begin{corollary}\label{cor:lowerbound} For all $t$,
\begin{eqnarray}
(i)\ R^{A}_{sum,t} &=& \min_{U^t} [I(X;U^t|Y)+I(Y;U^t|X)],
\label{eqn:minsumrate} \\
(ii)\ R^{A}_{sum,t} &\geq& H(f_B(X,Y)|Y)+H(f_A(X,Y)|X),
\label{eqn:lowerbound}
\end{eqnarray}
where in (i) $U^t$ are subject to all the Markov chain and
conditional entropy constraints in (\ref{eqn:rateregion}) and the
cardinality bounds given by (\ref{eqn:cardinality}).
\end{corollary}

{\em Proof:} For (i), add all the rate inequalities in
(\ref{eqn:rateregion}) enforcing all the constraints. Inequality (ii)
can be proved either using (\ref{eqn:minsumrate}) and relaxing the
Markov chains constraints, or using the following cut-set bound
argument. If $\mathbf Y$ is also available at location $A$, then
$\mathbf Z_B = f_B(\mathbf X, \mathbf Y)$ can be computed at location
$A$. Hence by the converse part of the Slepian-Wolf
theorem\cite{CoverThomas}, the sum-rate of all messages from $A$ to
$B$ must be at least $H(f_B(X,Y)|Y)$ for $B$ to form $\mathbf Z_B$.
Similarly, the sum-rate of all messages from $B$ to $A$ must be at
least $H(f_A(X,Y)|X)$. \hspace*{\fill}~\QED

Although (\ref{eqn:rateregion}) and (\ref{eqn:minsumrate}) provide
computable single-letter characterizations of ${\mathcal R}^A_t$ and
$R^A_{sum,t}$ respectively for all finite $t$, they do not provide a
characterization for $R_{sum,\infty}$ in terms of computable
single-letter information quantities. This is because the
cardinality bounds for the alphabets of the auxiliary random
variable $U^t$, given by (\ref{eqn:cardinality}), grow with $t$.

The Markov chain and conditional entropy constraints of
(\ref{eqn:rateregion}) imply certain structural properties which the
support-set of the joint distribution of the source and auxiliary
random variables need to satisfy. These properties are formalized
below in Lemma~\ref{lem:rectangle}. This lemma provides a bridge
between certain concepts which have played a key role in the
communication complexity literature\cite{CommComplexity} and
distributed source coding theory.  In order to state the lemma, we
need to introduce some terminology used in the communication
complexity literature\cite{CommComplexity}. This is adapted to our
framework and notation. A subset $\mathcal A \subseteq \mathcal X
\times \mathcal Y$ is called \emph{$f$-monochromatic} if the function
$f$ is constant on $\mathcal A$. A subset $\mathcal A \subseteq
\mathcal X \times \mathcal Y$ is called a \emph{rectangle} if
$\mathcal A=\mathcal S_X \times \mathcal S_Y$ for some $\mathcal S_X
\subseteq \mathcal X$ and some $\mathcal S_Y \subseteq \mathcal Y$.
Subsets of the form $\{x\} \times \mathcal{S}_Y$, $x \in
\mathcal{S}_X$, are called rows and subsets of the form $\mathcal S_X
\times \{y\}$, $y \in \mathcal{S}_Y$, are called columns the rectangle
$\mathcal A=\mathcal S_X \times \mathcal S_Y$. By definition, the
empty set is simultaneously a rectangle, a row, and a column. If each
row of a rectangle $\mathcal{A}$ is $f$-monochromatic, then
$\mathcal{A}$ is said to be row-wise $f$-monochromatic. Similarly, if
each column of a rectangle $\mathcal{A}$ is $f$-monochromatic, then
$\mathcal{A}$ is said to be column-wise $f$-monochromatic. Clearly, if
$\mathcal{A}$ is both row-wise and column-wise $f$-monochromatic, then
it is an $f$-monochromatic subset of $\mathcal X \times \mathcal Y$.

\begin{lemma}\label{lem:rectangle}
Let $U^t$ be any set of auxiliary random variables satisfying the
Markov chain and conditional entropy constraints of
(\ref{eqn:rateregion}). Let
$\mathcal{A}(u^t):=\{(x,y)|p_{XYU^t}(x,y,u^t)>0 \}$ denote the
projection of the $u^t$-slice of $\supp(p_{XYU^t})$ onto
$\mathcal{X}\times \mathcal{Y}$. If $\supp(p_{XY}) = \mathcal X \times
\mathcal Y$, then for all $u^t$, the following four conditions
hold. (i) $\mathcal{A}(u^t)$ is a rectangle. (ii) $\mathcal{A}(u^t)$
is row-wise $f_A$-monochromatic. (iii) $\mathcal{A}(u^t)$ is
column-wise $f_B$-monochromatic. (iv) If in addition, $f_A = f_B = f$,
then $\mathcal{A}(u^t)$ is $f$-monochromatic.
\end{lemma}

{\em Proof:} (i) The Markov chains in (\ref{eqn:rateregion}) induce
the following factorization of the joint pmf.
\begin{eqnarray*}\label{eqn:factorization}
p_{XYU^t}(x,y,u^t) &=& p_{XY}(x,y) \cdot p_{U_1|X}(u_1|x) \cdot
p_{U_2|YU_1}(u_2|y,u_1) \cdot \nonumber \\
&&p_{U_3|XU^2}(u_3|x,u^2)\ldots\nonumber\\
 &=:& p_{XY}(x,y) \phi_X(x,u^t) \phi_Y(y,u^t),
\end{eqnarray*}
where $\phi_X$ is the product of all the factors having conditioning
on $x$ and $\phi_Y$ is the product of all the factors having
conditioning on $y$. Let $\mathcal S_X(u^t):=\{x~|~\phi_X(x,u^t)>0\}$
and $\mathcal S_Y(u^t):=\{y~|~\phi_Y(y,u^t)>0\}$. Since
$p_{XY}(x,y)>0$ for all $x$ and $y$, $\mathcal A(u^t)=\mathcal
S_X(u^t)\times \mathcal S_Y(u^t)$. (ii) This follows from the
conditional entropy constraint $H(f_A(X,Y)|X,U^t)=0$ in
(\ref{eqn:rateregion}). (iii) This follows from the conditional
entropy constraint $H(f_B(X,Y)|Y,U^t)=0$ in (\ref{eqn:rateregion}).
(iv) This follows from parts (ii) and (iii) of this
lemma. \hspace*{\fill}~\QED

Note that $\mathcal A(u^t)$ is the empty set if, and only if,
$p_{U^t}(u^t)=0$. The above lemma holds for all values of $t$. The
fact that the set $\mathcal{A}(u^t)$ has a rectangular shape is a
consequence of the fact that the auxiliary random variables $U^t$ need
to satisfy the Markov chain constraints in
(\ref{eqn:rateregion}). These Markov chain constraints are in turn
consequences of the structural constraints which are inherent to the
coding process -- messages alternate from one terminal to the other
and can depend on only the source samples and all the previously
received messages which are available at a terminal. The rectangular
property depends ``less directly'' on the function-structure than on
the structure of the coding process and the structure of the joint
source distribution. On the other hand, the fact that
$\mathcal{A}(u^t)$ is row-wise or/and column-wise monochromatic is a
consequence of the fact that the auxiliary random variables $U^t$ need
to satisfy the conditional entropy constraints in
(\ref{eqn:rateregion}). This property is more closely tied to the
structure of the function and the structure of the joint distribution
of sources.  Lemma~\ref{lem:rectangle} will be used to prove
Theorems~\ref{thm:onereconnohelp} and Theorem~\ref{thm:infinitemsgLB}
in the sequel. \\

\subsection{Rate-distortion region} \label{sec:RD}

When per-sample distortion criteria are used, the single-letter
characterization of the rate-distortion region is given by
Theorem~\ref{thm:rateregion} with the conditional entropy constraints
in (\ref{eqn:rateregion}) replaced by the following expected
distortion constraints: there exist deterministic functions $\hat g_A$
and $\hat g_B$, such that $E\left[d_A(X,Y,\hat g_A(X,U^t))\right]\leq
D_A$ and $E\left[d_B(X,Y,\hat g_B(Y,U^t))\right]\leq D_B$. The proof
of achievability is similar to that of Theorem~\ref{thm:rateregion}.
The distortion constraints get satisfied ``automatically'' by using
strongly typical sets in the random coding and binning arguments.  The
proof of the converse given in Appendix~\ref{app:converse} will
continue to hold if equations (\ref{eqn:ZA}) and (\ref{eqn:ZB}) are
replaced by $E[d_A^{(n)}(\mathbf X, \mathbf Y, \widehat{\mathbf
    Z}_A)]\leq D_A + \epsilon$ and $E[d_B^{(n)}(\mathbf X, \mathbf Y,
  \widehat{\mathbf Z}_B)] \leq D_B+\epsilon$ respectively and the
subsequent steps in the proof changed appropriately.

The following proposition clarifies the relationship between the
rate region for probability of block error and the rate-distortion
region.

\begin{proposition}\label{prop:ratevsratedist}
Let $d_H$ denote the Hamming distortion function. If
$d_A(x,y,\hat{z}_A) = d_H(f_A(x,y),\hat{z}_A)$,
$d_B(x,y,\hat{z}_B)=d_H(f_B(x,y),\hat{z}_B)$, and $D_A=D_B=0$, then
$\{\mathbf{R}~|~ (\mathbf R, 0,0)\in
\mathcal{RD}^A_t\}=\mathcal{R}^A_t$.
\end{proposition}

{\em Proof:} In order to show that $\{\mathbf{R}~|~ (\mathbf R,
0,0)\in \mathcal{RD}^A_t\}\supseteq \mathcal{R}^A_t$, note that
$\forall \mathbf R\in \mathcal{R}^A_t$, we have $\epsilon \geq
\pP({\mathbf Z}_A \neq \widehat{\mathbf Z}_A)\geq
E[d_A^{(n)}(\mathbf X, \mathbf Y, \widehat{\mathbf Z}_A)]$ and
$\epsilon \geq \pP({\mathbf Z}_B \neq \widehat{\mathbf Z}_B)\geq
E[d_B^{(n)}(\mathbf X, \mathbf Y, \widehat{\mathbf Z}_B)]$ for the
distortion function assumed in the statement of the proposition.
Therefore $(\mathbf R,0,0) \in \mathcal{RD}^A_t$.

In order to show that $\{\mathbf{R}~|~ (\mathbf R, 0,0)\in
\mathcal{RD}^A_t\}\subseteq \mathcal{R}^A_t$, note that $\forall
\mathbf R$ such that $(\mathbf R,0,0) \in \mathcal{RD}^A_t$, we have
$d_A(X,Y,\hat g_A(X,U^t))=d_H(f_A(X,Y),\hat g_A(X,U^t))=0$, which
implies $f_A(X,Y)=\hat g_A(X,U^t)$, which in turn implies
$H(f_A(X,Y)|X,U^t)=0$. Similarly, we have $H(f_B(X,Y)|Y,U^t)=0$.
Therefore $\mathbf R\in \mathcal{R}^A_t$. \hspace*{\fill}~\QED

Although the proof of the single-letter characterization of
$\mathcal{RD}^A_t$ implies the proof of Theorem~\ref{thm:rateregion}
for $\mathcal{R}^A_t$, since the focus of this paper is on probability
of block error and the proofs for $\mathcal{RD}^A_t$ are very similar,
we provide the detailed converse proof only for
Theorem~\ref{thm:rateregion} for $\mathcal{R}^A_t$.

\section{Examples}\label{sec:examples}

Does interaction really help? In other words, does interactive coding
with more messages \emph{strictly} outperform coding with less
messages in terms of the sum-rate? When only one nontrivial function
has to be computed at only one location, at least one message is
needed. In this situation, interaction will be considered to be
``useful'' if there exists $t>1$ such that
$R_{sum,t}^A<R_{sum,1}^A$. When nontrivial functions have to be
computed at both locations, at least two messages are needed, one
going from $A$ to $B$ and the other from $B$ to $A$. Since messages go
in both directions, a two-message code can be potentially considered
to be interactive. However, this is a trivial form of interaction
because function computation is impossible without two
messages. Therefore, in this situation, interaction will be considered
to be useful if there exists $t>2$ such that
$R_{sum,t}^A<R_{sum,2}^A$.  Corollary~\ref{cor:lowerbound} does not
directly tell us if or when interaction is useful. In this section we
explore the value of interaction in different scenarios through some
striking examples.  Interaction does help in examples
\ref{subsec:functiononeloc}, \ref{subsec:DSBStwoloc} and
\ref{subsec:integralrate}, and does not (even with infinite messages)
in examples \ref{subsec:sourceoneloc}, \ref{subsec:sourcetwoloc} and
\ref{subsec:DSBSoneloc}.

\subsection{Interaction is useless for reproducing one source at one location: $f_A(x,y) := 0,
f_B(x,y) :=x $.}\label{subsec:sourceoneloc}

Only $\mathbf X$ needs to be reproduced at location $B$.  Unless
$H(X|Y)=0$, at least one message is necessary.  From
(\ref{eqn:lowerbound}), $\forall t \geq 1,~R^{A}_{sum,t} \geq
H(X|Y)$. But $R^A_{sum,1}=H(X|Y)$ by Slepian-Wolf coding
\cite{CoverThomas} with $\mathbf X$ as source and $\mathbf Y$ as
decoder side information. Hence, by Proposition~\ref{prop:struct}(i),
$R^{A}_{sum,t}=R^{A}_{sum,1}=H(X|Y)$ for all $t \geq 1$.

\subsection{Interaction is useless for reproducing both sources at
  both locations: $f_A(x,y):=y,
  f_B(x,y):=x$.}\label{subsec:sourcetwoloc}

Unless $H(X|Y)=0$ or $H(Y|X)=0$, at least two messages are
necessary. From (\ref{eqn:lowerbound}), $\forall t \geq
2,~R^{A}_{sum,t} \geq H(X|Y) + H(Y|X)$. But $R^A_{sum,2}=H(X|Y) +
H(Y|X)$ by Slepian-Wolf coding, first with $\mathbf X$ as source and
$\mathbf Y$ as decoder side information and then vice-versa. Hence, by
Proposition~\ref{prop:struct}(i), $R^{A}_{sum,t}=R^{A}_{sum,2}=H(X|Y)
+ H(Y|X)$ for all $t \geq 2$.

Examples~\ref{subsec:sourceoneloc} and \ref{subsec:sourcetwoloc}
show that if the goal is source reproduction with vanishing
distortion, interaction is useless\footnote{However, interaction can
prove useful for source reproduction when it is either required to
be error-free\cite{Orlitskyworst1,Orlitskyworst2} or when the
sources are stationary but non-ergodic\cite{Dakehe}}. To discover
the value of interaction, we must study either nonzero distortions
or functions which involve both sources. Our focus is on the latter.

\subsection{Benefit of interaction can be arbitrarily large for
function computation: $X \Perp Y$, $X \sim
\mbox{Uniform}\{1,\ldots,L\}$, $p_Y(1) = 1 - p_Y(0) = p\in(0,1)$,
$f_A(x,y):=0, f_B(x,y):=xy$ (real
multiplication).}\label{subsec:functiononeloc}

This is an expanded version of Example~8 in \cite{OrlitskyRoche}. At
least one message is necessary. If $t=1$, an achievable scheme is to
send $\mathbf X$ by Slepian-Wolf coding at the rate $H(X|Y)=\log_2 L$
so that the function can be computed at location $B$. Although
location $B$ is required to compute only the samplewise product and is
not required to reproduce ${\bf X}$, it turns out, rather
surprisingly, that the one-message rate $H(X|Y)$ cannot be
decreased. This is a direct consequence of a lemma due to Han and
Kobayashi which we now state by adapting it to our situation and
notation.

\begin{lemma}(Han and Kobayashi \cite[Lemma~1]{Han}) \label{lem:Han}
Let $\supp(p_{XY}) = \mathcal{X}\times \mathcal {Y}$. If $\forall x_1,
x_2\in \mathcal{X}$, $x_1\neq x_2$, there exists $y_0\in \mathcal {Y}$
such that $f_B(x_1,y_0)\neq f_B(x_2,y_0)$, then $R_{sum,1}^A\geq
H(X|Y)$.
\end{lemma}

The condition of Lemma~\ref{lem:Han} is satisfied in our present
example with $y_0 = 1$.  Therefore we have $R^A_{sum,1} =
H(X|Y)=\log_2 L$.  With one extra message and initial location $B$,
however, ${\mathbf Y}$ can be reproduced at location $A$ by
entropy-coding at the rate $R_1 = H(Y) = h_2(p)$ bits per
sample. Then, ${\mathbf Z}_B$ can be computed at location $A$ and
conveyed to location $B$ via Slepian-Wolf coding at the rate $R_2 =
H(f_B(X,Y)|Y) = p\log_2 L$ bits per sample, where $h_2$ is the binary
entropy function. Therefore, $R^B_{sum,2} \leq h_2(p)+p \log_2 L$. The
benefit of even one extra message can be significant: For fixed $L$,
$(R^A_{sum,1}/R^B_{sum,2})$ can be made arbitrarily large for suitably
small $p$. For fixed $p$, $(R^A_{sum,1} - R^B_{sum,2})$ can be made
arbitrarily large for suitably large $L$.

Extrapolating from this example, one might be led to believe that the
benefit of interaction arises due to computing nontrivial functions
which involve both sources as opposed to reproducing the sources
themselves. In other words, the function-structure determines whether
interaction is beneficial or not (recall that the sources were
independent in this example). However, the structure of the joint
distribution plays an equally important role and this aspect will be
highlighted in the next example.

\subsection{Interaction can be useless for computing any function at one location: $Y=X\oplus W$, $X\Perp W$, $X\sim
Ber(q)$, $W\sim Ber(p)$, $f_A(x,y):=0, f_B(x,y):=$ any function.
}\label{subsec:DSBSoneloc}

If $f_B(x,y)$ does not depend on $x$, i.e., there exists a function
$f'$ such that $f_B(x,y)=f'(y)$, no communication is needed and
interaction does not help.

If $f_B(x,y)$ depends on $x$, then $\exists~y_0\in\{0,1\}$ such that
$f_B(0,y_0)\neq f_B(1,y_0)$. Theorem~\ref{thm:onereconnohelp} below,
proved in Appendix~\ref{app:lemmaonerecon}, shows that interaction
does not help even with infinite messages.

\begin{theorem}\label{thm:onereconnohelp}
Let $f_A(x,y) =0$ and let $Y=X\oplus W$, with $X\Perp W$, $X\sim
Ber(q)$, and $W\sim Ber(p)$.  If there exists a $y_0\in\{0,1\}$ such
that $f_B(0,y_0)\neq f_B(1,y_0)$, then for all $t\in \zZ^+$,
$R^A_{sum,t}=H(X|Y)$.
\end{theorem}

{\em Remark:} The conclusion of Theorem~\ref{thm:onereconnohelp}
that interaction does not help \emph{cannot} be directly deduced
from (\ref{eqn:lowerbound}): When $f_B(x,y)=x\wedge y$ (Boolean
AND), the lower bound in Corollary~\ref{cor:lowerbound}(ii) $H(X
\wedge Y|Y)=H(X|Y=1)p_Y(1)$ is strictly less than $H(X|Y)$ if
$0<p,q<1$.

The result of Theorem~\ref{thm:onereconnohelp} can be generalized to
the following theorem for non-binary sources. The proof of this
theorem is provided in Appendix~\ref{app:lemmaonerecon} immediately
after the proof of Theorem~\ref{thm:onereconnohelp}.

\begin{theorem}\label{thm:nohelpgeneral}
Let $f_A(x,y) =0$ and let $\supp(p_{XY}) = \mathcal{X}\times
\mathcal{Y}$.  If (i) the only column-wise $f_B$-monochromatic
rectangles of $\mathcal{X}\times \mathcal{Y}$ are subsets of rows
and columns and (ii) there exists a random variable $W$ and
deterministic functions $\psi$ and $\eta$ such that $Y = \psi(X,W)$,
$X = \eta(Y,W)$, and $H(Y|X) = H(W)$,\footnote{It is easy to see
that if $Y = \psi(X,W)$, then $H(Y|X) = H(W)$ $\Leftrightarrow$
$X\Perp W$ and $H(W|X,Y) = 0$.} then for all $t\in \zZ^+$,
$R^A_{sum,t}=H(X|Y)$.
\end{theorem}

The examples till this point have highlighted the effects of
function-structure and distribution-structure on the benefit of
interaction. The next example will highlight a slightly different
aspect of function-structure associated with the situation in which
both sides need to compute the same nontrivial function which involves
both sources. The distribution-structure in the next example will be
essentially the same as in Example~\ref{subsec:DSBSoneloc} but with $q
= 1/2$ and $0<p<1$, i.e., $(X,Y) \sim$ DSBS$(p)$.  However, both
locations will need to compute the samplewise Boolean AND function.
Interestingly, in this situation the benefit of interaction returns as
explained below.

\subsection{Interaction can be useful for computing a function of sources at both locations:
$(X,Y) \sim$ DSBS$(p)$, $p \in (0,1)$, $f_A(x,y)=f_B(x,y) := x \wedge
  y$.}\label{subsec:DSBStwoloc}

Since both locations need to compute nontrivial functions, at least
two messages are needed. In a $2$-message code with initial location
$A$, location $B$ should be able to produce $\mathbf Z_B$ after
receiving the first message. By Lemma~\ref{lem:Han}, $R_1 \geq
H(X|Y)=h_2(p)$. With $R_1 =h_2(p)$ and a Slepian-Wolf code with
${\mathbf Y}$ as side-information, ${\mathbf X}$ can be reproduced at
location $B$. Thus for the second message, $R_2=H(f_B(X,Y)|X)=(1/2)
h_2(p)$ is both necessary and sufficient to ensure that location $A$
can produce ${\mathbf Z}_A$.  Hence $R^A_{sum,2}=(3/2) h_2(p)$.

If a third message is allowed, one choice of auxiliary random
variables in (\ref{eqn:rateregion}) is $U_1:=X \vee W$, $W\sim
Ber(1/2), W \Perp (X,Y)$, $U_2:=Y \wedge U_1$, and $U_3:=X \wedge
U_2$. Hence $U_3 = X \wedge Y=f_B(X,Y) \Rightarrow
H(f_A(X,Y)|X,U^3)=H(f_B(X,Y)|Y,U^3)=0$. Hence, $ R^A_{sum,3} \leq
I(X;U^3|Y)+I(Y;U^3|X) =\frac{5}{4} h_2(p)+\frac{1}{2}
h_2\left(\frac{1-p}{2}\right) - \frac{(1-p)}{2} \stackrel{(a)}{<}
\frac{3}{2} h_2(p) = R^A_{sum,2}$, where step $(a)$ holds for all $p
\in (0,1)$ and the gap is maximum for $p = 1/3$. When $p=0.5$, $X
\Perp Y$, and an achievable $3$-message sum-rate is $\approx 1.406 <
1.5 = R^A_{sum,2}$.

Note that as a special case of Example~\ref{subsec:DSBSoneloc}, if
$(X,Y) \sim$ DSBS$(p)$ and only location $B$ needs to compute the
Boolean AND function, interaction is useless. But if both locations
need to compute it, and $p \in (0,1)$, then the benefit of interaction
returns. Motivated by the benefits of using the more and more
messages, we investigate infinite-message interaction in the following
example.

\subsection{An achievable infinite-message sum-rate as a definite
integral with infinitesimal-rate messages: $X \Perp Y$, $X \sim
Ber(p)$, $Y \sim Ber(q)$, $p, q \in (0,1)$, $f_A(x,y) = f_B(x,y) = x
\wedge y$.}\label{subsec:integralrate}
\begin{figure*}[!htb]
\begin{center}
\scalebox{0.5}{\input{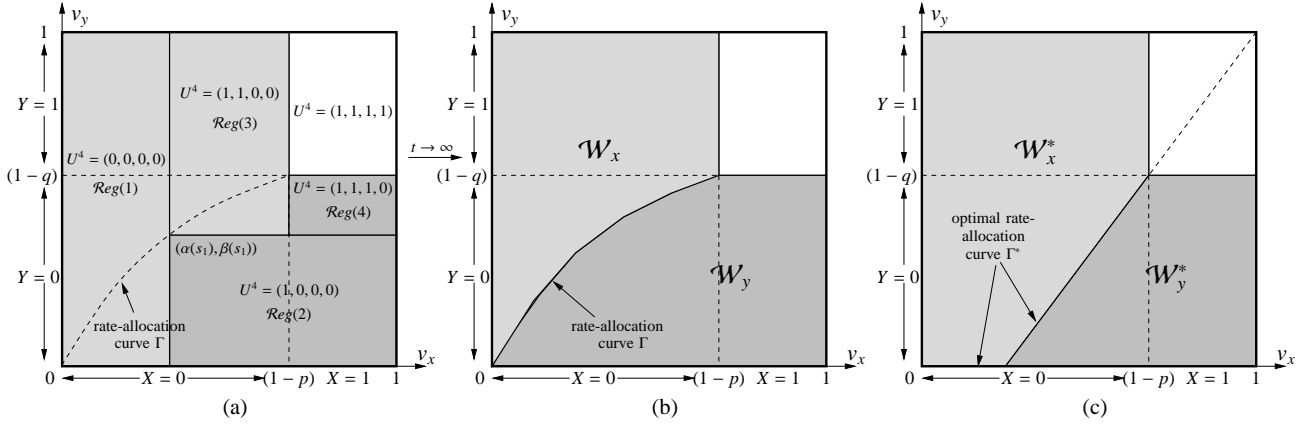}}
\end{center}
\caption{\sl (a) $4$-message interactive code (b) $\infty$-message
interactive code (c) $\infty$-message interactive code with optimal
rate-allocation curve when $q \geq p$. \label{fig:integral}}
\end{figure*}

As in Example~\ref{subsec:DSBStwoloc}, the $2$-message minimum
sum-rate is $R^A_{sum,2}=H(X|Y)+H(f_B(X,Y)|X)=h_2(p)+p h_2(q)$.
Example~\ref{subsec:DSBStwoloc} demonstrates the gain of
interaction. This inspires us to generalize the $3$-message code of
Example~\ref{subsec:DSBStwoloc} to an arbitrary number of messages
and evaluate an achievable infinite-message sum-rate. Since we are
interested in the limit $t \rightarrow \infty$, it is sufficient to
consider even-valued $t$ due to Proposition~\ref{prop:struct}.

Define real auxiliary random variables $(V_x,V_y) \sim
\mbox{Uniform}([0,1]^2)$. If $X:=1_{[1-p,1]}(V_x)$ and
$Y:=1_{[1-q,1]}(V_y)$, then $(X,Y)$ has the correct joint pmf, i.e.,
$p_X(1)=1-p_X(0)=p$, $p_Y(1)=1-p_Y(0)=q$ and $X\Perp Y$. We will
interpret $0$ and $1$ as real zero and real one respectively as
needed. This interpretation will allow us to express Boolean
arithmetic in terms of real arithmetic. Thus $X \wedge Y$ (Boolean
AND) $= XY$ (real multiplication). Define a {\em rate-allocation
  curve} $\Gamma$ parametrically by $\Gamma := \{(\alpha(s),
\beta(s)), 0 \leq s \leq 1\}$ where $\alpha$ and $\beta$ are real,
nondecreasing, absolutely continuous functions with
$\alpha(0)=\beta(0)=0$, $\alpha(1)=(1-p)$, and $\beta(1)=(1-q)$. The
significance of $\Gamma$ will become clear later.  Now choose a
partition of $[0,1]$, $0 = s_0 < s_1 <\ldots <s_{t/2-1}<s_{t/2}=1$,
such that $\max_{i=1,\ldots,t/2}(s_i-s_{i-1})<\Delta_t$.  For
$i=1,\ldots,t/2$, define $t$ auxiliary random variables as follows,
\[
U_{2i-1}:= 1 _{[\alpha(s_i),1]\times [\beta(s_{i-1}),1]}(V_x,V_y),~
U_{2i}:= 1_{[\alpha(s_i),1]\times [\beta(s_i),1]}(V_x,V_y).
\]

In Figure~\ref{fig:integral}(a), $(V_x,V_y)$ is uniformly distributed
on the unit square and $U^t$ are defined to be $1$ in rectangular
regions which are nested.  The following properties can be verified:

\begin{itemize}
\item[$P1$:] $U_1 \geq U_2 \geq \ldots \geq U_t$.
\item[$P2$:] $H(X \wedge Y| X, U^t)=H(X \wedge Y| Y, U^t)=0$: since
$U_{t}=1_{[1-p,1] \times [1-q,1]}(V_x,V_y)=X \wedge Y$.
\item[$P3$:] $U^t$ satisfy all the Markov chain constraints in
(\ref{eqn:rateregion}): for example, consider $U_{2i} - (Y,U^{2i-1})-
X$.  $U_{2i-1}=0 \Rightarrow U_{2i}=0$ and the Markov chain
holds. $U_{2i-1}= Y =1 \Rightarrow (V_x, V_y) \in [\alpha(s_i),1]
\times [1-q,1] \Rightarrow U_{2i}=1$ and the Markov chain holds. Given
$U_{2i-1}=1, Y=0$, $(V_x,V_y) \sim \mbox{Uniform}([\alpha(s_i),1]
\times [\beta(s_{i-1}),1-q]) \Rightarrow$ $V_x $ and $V_y$ are
conditionally independent. Thus $X \Perp U_{2i}| (U_{2i-1}=1, Y=0)$
because $X$ is a function of only $V_x$ and $U_{2i}$ is a function of
only $V_y$ upon conditioning. So the Markov chain $U_{2i} -
(Y,U^{2i-1})- X$ holds in all situations.
 \item[$P4$:] $(Y,U_{2i}) \Perp X |U_{2i-1}=1$: this can be proved by
 the same method as in $P3$.
\end{itemize}
$P2$ and $P3$ show that $U^t$ satisfy all the constraints in
(\ref{eqn:rateregion}).

For $i = 1,\ldots,t/2$, the $(2i)$-th rate is given by
\begin{eqnarray*}
\lefteqn{I(Y;U_{2i}|X,U^{2i-1}) =} &&\\
&\stackrel{P1}{=}&I(Y;U_{2i}|X,U_{2i-1}=1)p_{U_{2i-1}}(1)\\
&\stackrel{P4}{=}&I(Y;U_{2i}|U_{2i-1}=1)p_{U_{2i-1}}(1)\\
&=&H(Y|U_{2i-1}=1)p_{U_{2i-1}}(1)-H(Y|U_{2i},
U_{2i-1}=1)p_{U_{2i-1}}(1)
\\
&\stackrel{(b)}{=}&H(Y|U_{2i-1}=1)p_{U_{2i-1}}(1)-H(Y|U_{2i}=1)p_{U_{2i}}(1)\\
&=&
(1-\alpha(s_i))\left((1-\beta(s_{i-1}))h_2\left(\frac{q}{1-\beta(s_{i-1})}\right)
\right.\\ &&\left.
-(1-\beta(s_i))h_2\left(\frac{q}{1-\beta(s_i)}\right)\right)
\\
&\stackrel{(c)}{=}& (1-\alpha(s_i)) \int_{\beta(s_{i-1})}^{\beta(s_i)} \log_2 \left(\frac{1-v_y}{1-q-v_y}\right) d v_y \\
&=&
\int\!\!\!\!\int_{[\alpha(s_i),1]\times[\beta(s_{i-1}),\beta(s_i)]}
w_y(v_y,q) d v_x d v_y,
\end{eqnarray*}
where step (b) is due to property $P4$ and because
$(U_{2i-1},U_{2i})=(1,0) \Rightarrow Y=0$, hence $H(Y|U_{2i},
U_{2i-1}=1)p_{U_{2i-1}}(1) =
H(Y|U_{2i}=1,U_{2i-1}=1)p_{U_{2i},U_{2i-1}}(1,1)\stackrel{P1}{=}H(Y|U_{2i}=1)p_{U_{2i}}(1)$,
and step (c) is because
\[
\frac{\partial}{\partial v_y} \left( -(1-v_y)h_2\left(
\frac{q}{1-v_y}\right) \right) = \log_2 \left(\frac{1-v_y}{1-q-v_y}\right)=:
w_y(v_y,q).
\]
The $2i$-th rate can thus be expressed as a 2-D integral of a weight
function $w_y$ over the rectangular region ${\mathcal
  Reg}(2i):=[\alpha(s_i),1]\times[\beta(s_{i-1}),\beta(s_i)]$ (a
horizontal bar in Figure~\ref{fig:integral}(a)). Therefore, the sum of
rates of all messages sent from location $B$ to location $A$ is the
integral of $w_y$ over the union of all the corresponding horizontal
bars in Figure~\ref{fig:integral}(a). Similarly, the sum of rates of
all messages sent from location $A$ to location $B$ can be expressed
as the integral of another weight function
$w_x(v_x,p):=\log_2((1-v_x)/(1-p-v_x))$ over the union of all the
vertical bars in Figure~\ref{fig:integral}(a).

Now let $t\rightarrow \infty$ such that $\Delta_t \rightarrow 0$.
Since $\alpha$ and $\beta$ are absolutely continuous,
$(\alpha(s_i)-\alpha(s_{i-1}))\rightarrow 0$ and
$(\beta(s_i)-\beta(s_{i-1}))\rightarrow 0$.  The union of the
horizontal (resp.~vertical bars) in Figure~\ref{fig:integral}(a) tends
to the region ${\mathcal W}_y$ (resp.~${\mathcal W}_x$) in
Figure~\ref{fig:integral}(b).  Hence an achievable infinite-message
sum-rate given by
\begin{equation}
\int\!\!\!\!\int_{{\mathcal W}_x} w_x(v_x,p) dv_x
dv_y+\int\!\!\!\!\int_{{\mathcal W}_y} w_y(v_y,q) dv_x
dv_y\label{eqn:integralrate}
\end{equation}
depends on only the rate-allocation curve $\Gamma$ which coordinates
the progress of source descriptions at $A$ and $B$. Since ${\mathcal
  W}_x \bigcup {\mathcal W}_y$ is independent of $\Gamma$,
(\ref{eqn:integralrate}) is minimized when $\mathcal{W}_x =
\mathcal{W}_x^* := \{(v_x,v_y) \in [0,1-p] \times [0,1-q]: w_x(v_x,p)
\leq w_y(v_y,q)\} \cup [0,1-p]\times[1-q,1]$. For $q \geq p$, the
boundary $\Gamma^*$ separating $\mathcal{W}_x^*$ and $\mathcal{W}_y^*$
is given by the piecewise linear curve connecting $(0,0)$,
$((q-p)/q,0)$, $(1-p,1-q)$ in that order (see Figure~2(c)).

For $\mathcal{W}_x = \mathcal{W}_x^*$, (\ref{eqn:integralrate}) can be
evaluated in closed form and is given by
\begin{equation}
h_2(p)+p h_2(q)+p \log_2 q + p(1-q) \log_2
e.\label{eqn:infmsgsumrate}
\end{equation}
Recall that $R^A_{sum,2}=h_2(p)+p h_2(q)$. The difference $p(\log_2
q+(1-q) \log_2 e)$ is an increasing function of $q$ for $q\in (0,1]$
and equals $0$ when $q=1$. Hence the difference is negative for $q\in
(0,1)$. So $R_{sum,\infty} < R^A_{sum,2}$ and interaction does
help. In particular, when $p=q=1/2$, ($(X,Y) \sim$ iid $Ber(1/2)$), by
an infinite-message code, we can achieve the sum-rate $(1+(\log_2
e)/4)\approx 1.361$, compared with the $3$-message achievable sum-rate
$1.406$ and the $2$-message minimum sum-rate $1.5$ in
Example~\ref{subsec:DSBStwoloc}. It should be noted that for finite
$t$, $\Gamma$ is staircase-like and contains horizontal and vertical
segments. However, $\Gamma^*$ contains an oblique segment. So the code
with finite $t$ generated in this way never achieves the
infinite-message sum-rate. It can be approximated only when
$t\rightarrow \infty$ and each message uses an infinitesimal rate.

Note that the achievable sum-rate (\ref{eqn:integralrate}) is not
shown to be the optimal sum-rate $R_{sum,\infty}$ because we only
consider a particular construction of the auxiliary random
variables. We have, however, the following lower bound for
$R_{sum,\infty}$ which can be proved by a technique which is similar
to the proof of Theorem~\ref{thm:onereconnohelp}.

\begin{theorem}\label{thm:infinitemsgLB}
If $X\Perp Y, X \sim Ber(p), Y \sim Ber(q), f_A(x,y)=f_B(x,y)= x
\wedge y$, $0<p,q<1$, we have
\[R_{sum,\infty} \geq h_2(p)+h_2(q)-(1-pq) h_2\left(\frac{(1-p)(1-q)}{1-pq}\right).\]
\end{theorem}

The proof is given in Appendix~\ref{app:LBinfmsg}. This lower bound is
strictly less than (\ref{eqn:infmsgsumrate}) when $0<p,q<1$. For
example, when $p=q=1/2$, ($(X,Y)\sim \mbox{iid } Ber(1/2)$), the bound
in Theorem~\ref{thm:infinitemsgLB} gives us $R_{sum,\infty}\geq
(2-(3/4)h_2(1/3)) \approx 1.311$, compared with the infinite-message
achievable sum-rate $1.361$.

\section{Multiterminal interactive function computation\label{sec:multiterminal}}

We can consider multiterminal interactive function computation
problems as generalizations of the two-terminal interactive function
computation problem.  At a high level, interactive function
computation may be thought of as a form of distributed source coding
with progressive levels of feedback. Although the multiterminal
problem is significantly more intricate, important insights can be
extracted by leveraging results for the two-terminal problem.  The
ability to progressively refine information bi-directionally in
multiple rounds
lies at the heart of interactive function computation.
This ability to refine information can have a significant impact on
the efficiency of information transport in large networks as discussed
in Section~\ref{subsec:multiexamples} (see Example~3).

\subsection{\label{subsec:multiproblem}Problem formulation}

Let $m$ be the number of nodes. Consider $m$ statistically dependent
discrete memoryless stationary sources taking values in finite
alphabets.  For each $j$, where $j$ takes integer values $1$ through
$m$, let $\mathbf X_j:=(X_j(1),\ldots,X_j(n))\in (\mathcal{X}_j)^n$
denote the $n$ source samples which are available at node $j$.  For
$i=1,\ldots,n$, let $(X_1(i),X_2(i),\ldots,X_m(i))\sim$ iid
$p_{X_1,\ldots,X_m}$ where $p_{X_1,\ldots,X_m}$ is a joint pmf which
describes the statistical dependencies among the samples observed at
the $m$ nodes at each time instant. For each $j$ and $i$, let
$Z_j(i):=f_j(X_1(i),\ldots,X_m(i)) \in \mathcal{Z}_j$ and let $\mathbf
Z_j := (Z_j(1),\ldots,Z_j(n))$. The tuple $\mathbf Z_j$ denotes $n$
samples of the samplewise function of all the sources which is desired
to be computed at node $j$.

Let the topology of the network be characterized by a directed graph
$\mathcal G=(\mathcal V,\mathcal E)$, where $\mathcal
V:=\{1,\ldots,m\}$ is the vertex set of all the nodes and $\mathcal E$
is the edge set of all the directed links which are available for
communication. The network topology describes the connectivity and
information flow constraints in the network.
It is assumed that the topology is consistent with the goals of
function computation, that is, for every node which computes a
nontrivial function which depends on the source samples at other
nodes, there exists a set of directed paths over which information can
be transfered from the relevant nodes to perform the computation. In
order to perform the computations, a $t$-round multiterminal
interactive distributed source code for function computation can be
defined by extending the notion of a $t$-round concurrent-message
interactive code for the two-terminal problem (see
Section~\ref{subsec:concurrent}) in the following manner. In the
$i$-th round, where $i$ takes integer values $1$ through $t$, for each
directed link $(j,k)\in \mathcal E$, a message $M_{jki}$ is generated
at node $j$ as a pre-specified deterministic function of $\mathbf X_j$
and all the messages to and from this node in all the previous rounds.
Then all the messages in the $i$-th round are transferred concurrently
over all the available directed links.  After $t$ rounds, at each node
$j$, a decoding function reproduces $\mathbf Z_j$ as $\widehat
{\mathbf Z}_j$ based on $\mathbf X_j$ and all the messages to and from
this node. As part of the $t$-round interactive code specification, a
message over any link in any round is allowed to be a null message,
i.e, no message is sent over the link, and this is known in advance as
part of the code.  By incorporating null messages, the
concurrent-message interactive coding scheme described above subsumes
all conceivable types of interaction. Let a link be called {\em
active} in a given round if it does not carry a null message in that
round.  For each round $i$, let $\mathcal{E}_i$ denote the subset of
directed links in $\mathcal{E}$ which are active. A $t$-round
interaction protocol is the sequence of directed subgraphs
$\mathcal{E}_1,\ldots,\mathcal{E}_t$ which describes how the nodes are
permitted to exchange messages over different rounds. This controls
the dynamics of information flow in the network.

\begin{figure*}[!htb]
\begin{center}
\scalebox{0.5}{\input{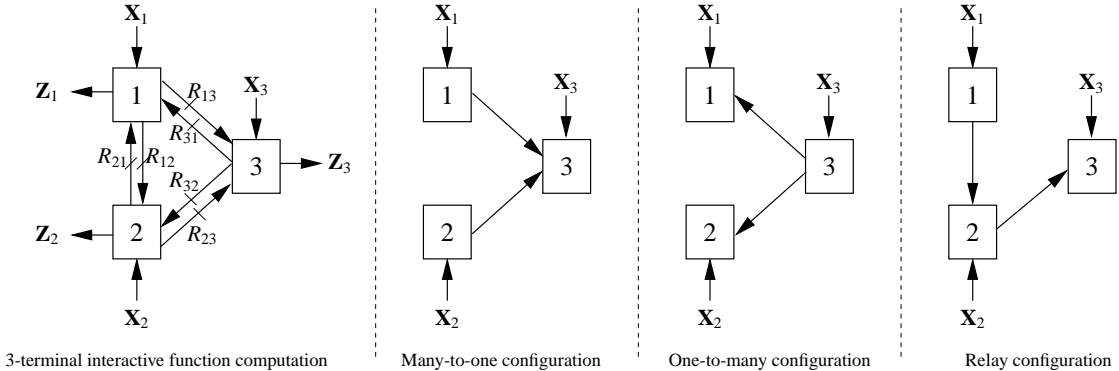}}
\end{center}
\caption{\label{fig:modes} \small \sl Interactive function computation
  can be viewed as an interaction protocol which successively switches
  among several basic distributed source coding configurations.
}
\end{figure*}

Our key point of view, illustrated in Figure~\ref{fig:modes}, is that,
interactive function computation is at its heart, an interaction
protocol which successively switches the information-flow topology
among several basic distributed source coding configurations. In the
two-terminal case, the alternating-message interaction protocol is
simple: messages alternate from one node to the other; the only free
parameter in the protocol being the initial node which must be chosen
to minimize the sum rate.  For this protocol, there is essentially
only one type of configuration and accordingly only one basic
distributed source coding strategy, namely, Wyner-Ziv-like coding with
all the previously received messages as common side-information
available to both the nodes.  The multiterminal case is, however,
significantly more intricate. For instance, with three nodes there are
several basic configurations in addition to the point-to-point one,
e.g., many-to-one, one-to-many, and relay as shown in
Figure~\ref{fig:modes}.

The efficiency of communication for function computation can be
measured at various levels.  The most precise characterization would
be in terms of
the $(t|\mathcal E|)$-dimensional rate tuple $(R_{jki})_{(j,k)\in
\mathcal E,i=1,\ldots,t}$ corresponding to the number of bits per
sample in \emph{each link} in \emph{each round}.  A coarser
characterization would be in terms of
the $|\mathcal E|$-dimensional total-rate tuple $(R_{jk})_{(j,k) \in
\mathcal{E}}$, where $R_{jk}$ is the total number of bits per sample
transferred through link $(j,k)$ in \emph{all the rounds}. The
coarsest characterization would be in terms of the sum-total-rate
which is the sum of the total number of bits per sample in \emph{all
the rounds} through \emph{all the links}. One can then define
admissible rates, admissible total-rates, and the minimum
sum-total-rate $R_{sum,t}$, following Definition~\ref{def:rateregion},
in terms of rates for which there exist encoding and decoding
functions for which the block error probability of function
computation goes to zero as the blocklength goes to infinity.  Let
$t^*$ denote the minimum number of rounds for which function
computation is feasible. Computation is nontrivial if $t^* \geq 1$.
Clearly, $t^*$ is not more than the diameter of the largest connected
component of the network which is itself not more than $(m-1)$. Hence
$t^* \leq (m-1)$.  We will consider interaction to be useful if
$R_{sum,t} < R_{sum,t^*}$ for some $t > t^*$.

The search for an optimum interactive code is a twofold search over
all interaction protocols and over all distributed source codes. The
interaction protocol dictates which nodes transmit and which nodes
receive messages in each round.  The distributed source code dictates
what information to send and how to decode it.  In the two-terminal
case, the standard machinery of random coding and binning is adequate
to characterize the rate region and the minimum sum rate because it
can be viewed as a sequence of Wyner-Ziv-like codes. In the
multiterminal case, however, finding a computable characterization of
the rate regions in terms of single-letter information measures can be
challenging because the rate regions for even non-interactive special
cases, such as the many-to-one, one-to-many, and relay configurations
(see Figure~\ref{fig:modes}) are longstanding open problems. For many
of these configurations, the standard machinery of random coding and
binning fall short of giving the optimal performance as exemplified by
the K\"{o}rner-Marton problem \cite{KornerMarton}. These difficulties
notwithstanding, results for the two-terminal interactive function
computation problem can be used to develop insightful performance
bounds and architectural guidelines for the general multiterminal
problems. This is discussed in the following two subsections.

\subsection{\label{subsec:cutset}Cut-set bounds} \label{subsec:cutsetbnd}

Given any $t$-round multiterminal interactive function computation
problem, we can formulate a $t$-round two-terminal interactive
function computation problem with concurrent messages by regarding a
set of nodes $\mathcal S \subseteq \mathcal V$ as one terminal and the
complement $\mathcal S^c$ as the other. The minimum sum-rate for this
two-terminal problem is a lower bound for the minimum sum-total-rate
between $\mathcal S$ and $\mathcal {S}^c$ in the original
multiterminal problem.

Let $R_{A, B}:=\sum_{j\in A, k\in B, (j,k)\in\mathcal{E}}R_{jk}$
denote the sum-total-rate from a set of nodes $A$ to a set of nodes
$B$ (over all rounds and over all available directed links from $A$ to
$B$). Let $R_{sum,t}^{\mathcal S, \mathcal S^c}$ denote the minimum
sum-rate of the $t$-round two-terminal problem with concurrent
messages with sources $(X_j)_{j\in \mathcal {S}}$ at $A$ and
$(X_j)_{j\in \mathcal {S}^c}$ at $B$ and functions $(f_j(X^m))_{j\in
\mathcal {S}}$ and $(f_j(X^m))_{j\in \mathcal {S}^c}$ to be computed
at $A$ and $B$ respectively. A systematic method for developing
cut-set lower bounds for the minimum sum-total-rate of the $t$-round
multiterminal problem is to formulate a linear program with
$(R_{jk})_{(j,k)\in\mathcal{E}}$ as the variables and the
sum-total-rate $\sum_{(j,k)\in\mathcal{E}} R_{jk}$ as the linear
objective function to be minimized subject to the following linear
inequality constraints: $\forall \mathcal{S}\subseteq \mathcal{V}$,
$R_{\mathcal S, \mathcal S^c}\geq H((f_j(X^m))_{j\in \mathcal {S}^c}|
(X_j)_{j\in \mathcal {S}^c})$, $R_{\mathcal S^c, \mathcal S}\geq
H((f_j(X^m))_{j\in \mathcal {S}}| (X_j)_{j\in \mathcal {S}})$,
$(R_{\mathcal S, \mathcal S^c}+R_{\mathcal S^c, \mathcal S}) \geq
R_{sum,t}^{\mathcal S, \mathcal S^c}$, and $R_{jk}\geq 0, \forall
j\neq k$. Note that the first two constraints respectively come from
the first two terms on the right side of
Corollary~\ref{cor:lowerbound}~(ii). Such cut-set bounds can often
provide insights into when interaction may be useful and when it may
not be (see examples below).

\subsection{\label{subsec:multiexamples}Examples}

\noindent{\em Example~1:} Consider three nodes with sources
$(X_1,X_2)\sim$ DSBS$(p)$, $p \in (0,1)$, and $X_3=0$. The functions
desired at nodes $1$, $2$, and $3$ are $f_1= 0$, $f_2=0$, and
$f_3(x_1,x_2)=x_1\oplus x_2$ respectively. In other words, correlated
sources $\mathbf X_1$ and $\mathbf X_2$ are available at nodes~1 and 2
respectively, and node~3 needs to compute the samplewise Boolean XOR
function $\mathbf X_1 \oplus \mathbf X_2$.  Assume that this
three-terminal network has a fully connected topology $\mathcal{E}$.
\begin{figure}[!htb]
\begin{center}
\scalebox{0.5}{\input{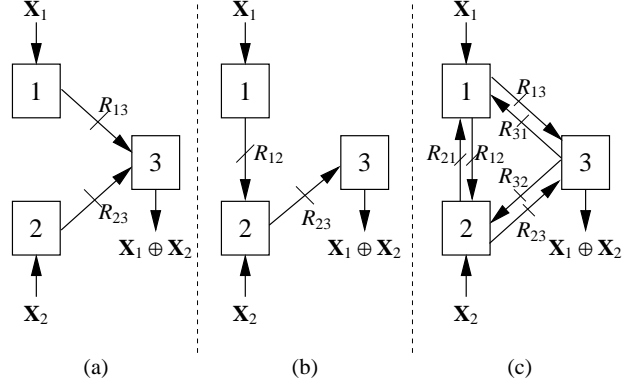}}
\end{center}
\caption{\sl (a) Many-to-one K\"{o}rner-Marton scheme. (b) Relay
  scheme. (c) General interactive scheme. When $(X_1,X_2)\sim$
  DSBS$(p)$, $p \in (0,1)$, all three schemes have the same minimum
  sum-total-rate $2 h_2(p)$. \label{fig:kornermarton}}
\end{figure}

First consider the $1$-round many-to-one interaction protocol given by
$\mathcal E_1=\{(1,3),(2,3)\}$.  Under this interaction protocol, the
distributed function computation problem reduces to the
K\"{o}rner-Marton problem \cite{KornerMarton} and is illustrated in
Figure~\ref{fig:kornermarton}(a). The distributed source coding scheme
of K\"{o}rner and Marton based on binary linear codes (see
\cite{KornerMarton}) achieves the goal of computing the Boolean XOR at
node $3$ with $R_{131} = R_{231} = R_{13} = R_{23} = H(X_1\oplus X_2)
= h_2(p)$. Hence the sum-total-rate of this non-interactive
many-to-one coding scheme is given by $R_{13} + R_{23} = 2 h_2(p)$
bits per sample. Thus, in this example $t^* = 1$ and the coding is
non-interactive.

Next, consider the $2$-round relay-based interaction protocol given by
$\mathcal E_1=\{(1,2)\}$ and $\mathcal E_2=\{(2,3)\}$ as illustrated
in Figure~\ref{fig:kornermarton}(b).  Consider the following coding
strategy.  Using Slepian-Wolf coding in the first round, with $R_{121}
= R_{12} = H(X_1|X_2)=h_2(p)$, ${\mathbf X}_1$ can be reproduced at
node $2$. Then, $\mathbf X_1 \oplus \mathbf X_2$ can be computed at
node $2$ and the result of the computation can be conveyed to node $3$
in the second round by entropy-coding at the rate given by $R_{232} =
R_{23} = H(X_1\oplus X_2)=h_2(p)$.  Hence the sum-total-rate of this
relay scheme is given by $R_{12}+R_{23}=2 h_2(p)$ bits per sample.
Since under this protocol information is constrained to flow in only
one direction from source node $1$ to source node $2$ in round one and
then from node $2$ to the destination node $3$ in round two,
distributed source codes which respect this protocol are, truly
speaking, non-interactive.

Finally, consider general $t$-round interactive codes. The cut-set
lower bound between $\{1\}$ and $\{2,3\}$ for computing $\mathbf X_1
\oplus \mathbf X_2$ at $\{2,3\}$ gives $ R_{12} + R_{13} \geq
H(X_1\oplus X_2|X_2) = h_2(p)$. Interchanging the roles of nodes $1$
and $2$ in the previous cut-set bound we also have $ R_{21} + R_{23}
\geq H(X_1\oplus X_2|X_1) = h_2(p)$. Adding these two bounds gives
$R_{12} + R_{13} + R_{21} + R_{23} \geq 2 h_2(p)$. Hence, $R_{sum,t}
\geq 2 h_2(p)$. This shows that the sum-total-rates of the many-to-one
K\"{o}rner-Marton and the relay schemes are optimum. No amount of
interaction can reduce the sum-total-rate of these non-interactive
schemes.  \\

\noindent{\em Example~2:} Consider three nodes with sources
$(X_1,X_2)\sim$ DSBS$(p)$, $p \in (0,1)$, and $X_3=0$. The functions
desired at nodes $1$, $2$, and $3$ are $f_1= 0$, $f_2=0$, and
$f_3(x_1,x_2)=x_1\wedge x_2$ respectively. In other words, correlated
sources $\mathbf X_1$ and $\mathbf X_2$ are available at nodes~1 and 2
respectively, and node~3 needs to compute the samplewise Boolean AND
function instead of the XOR function in Example~1. As in Example~1,
assume that this three-terminal network has a fully connected topology
$\mathcal{E}$.
\begin{figure}[!htb]
\begin{center}
\scalebox{0.5}{\input{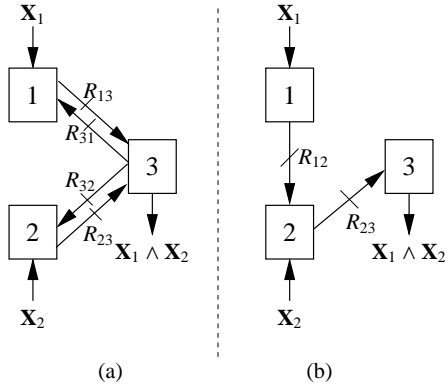}}
\end{center}
\caption{\sl (a) Interactive Many-to-one scheme. (b) Relay scheme.
  When $(X_1,X_2)\sim$ DSBS$(p)$ and $p \in (1/3, 1)$, the minimum
  sum-total-rate for (b) is less than that for
  (a). \label{fig:threetermAND}}
\end{figure}

Consider a general $t$-round interactive code with the following
interaction protocol: for all $i = 1, \ldots, t$, $\mathcal{E}_i = \{
(1,3), (3,1), (2,3), (3,2)\}$ (see Figure~\ref{fig:threetermAND}(a)).
Note that nodes~$1$ and $2$ cannot directly communicate with each
other under this interaction protocol. Due to
Theorem~\ref{thm:onereconnohelp}, the cut-set lower bound between
$\{1\}$ and $\{2,3\}$ for computing $\mathbf X_1 \wedge \mathbf X_2$
at $\{2,3\}$ is given by: $R_{13} + R_{31} \geq H(X_1|X_2) =
h_2(p)$. Similarly, we have $ R_{23} + R_{32} \geq H(X_2|X_1) =
h_2(p)$. Adding these two bounds gives $R_{13} + R_{31} + R_{23} +
R_{32} \geq 2 h_2(p)$. It should be clear that $t^* = 1$ because nodes
$1$ and $2$ can send all their source samples to node $3$ in one
round. If there is only one round, there is no advantage to be gained
by transferring messages between nodes $1$ and $2$. This observation,
together with the above cut-set bound shows that $R_{sum,t^*} \geq 2
h_2(p)$.

Now consider the $2$-round relay scheme illustrated in
Figure~\ref{fig:threetermAND}(b).  Using Slepian-Wolf coding in the
first round, with $R_{121} = R_{12} = H(X_1|X_2)=h_2(p)$, ${\mathbf
  X}_1$ can be reproduced at node~$2$. Then, $\mathbf X_1 \wedge
\mathbf X_2$ can be computed at node~$2$ and the result of the
computation can be conveyed to node~$3$ in the second round by
entropy-coding at the rate given by $R_{232} = R_{23} = H(X_1\wedge
X_2)= h_2\left(\frac{1-p}{2}\right)$. Hence the sum-total-rate of this
relay scheme is given by $R_{12} + R_{23} = h_2(p) +
h_2\left(\frac{1-p}{2}\right)$ bits per sample, which is less than $2
h_2(p)$ when $p > 1/3$. Thus, for $p > 1/3$, $R_{sum,2} < R_{sum,t^*}$
and interaction is useful.\footnote{Truly speaking, this coding scheme
  is non-interactive because information flows in only one direction
  from node $1$ to node $2$ and then from node $2$ to node $3$.} In
fact, when $p > 1/3$, \emph{a single message} from node $1$ to node
$2$ is more beneficial in terms of the sum-total-rate than multiple
rounds of two-way communication between nodes $1$ and $3$ and between
nodes $2$ and $3$. \\

\begin{figure}[!htb]
\begin{center}
\scalebox{0.5}{\input{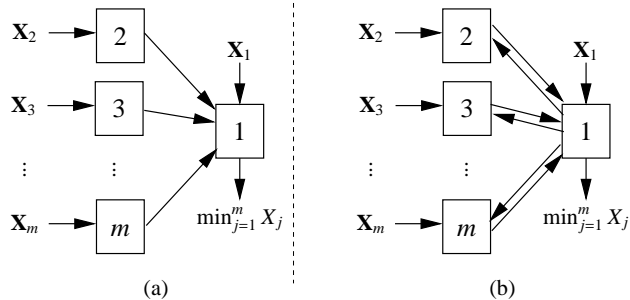}}
\end{center}
\caption{\sl (a) Non-interactive function computation. (b)
Interactive function computation. When $(X_1,\ldots,X_m)\sim$ iid
$Ber(1/2)$, the minimum sum-total-rate for (b) is orderwise smaller
than that for (a). \label{fig:starnetwork}}
\end{figure}

\noindent{\em Example~3:} Consider $m \geq 3$ nodes and $m$
independent sources $X_1, \ldots, X_m$ each of which is iid
$Ber(1/2)$.  For each $i$, the $i$-th source $X_i$ is observed at only
the $i$-th node. Only node $1$ needs to compute the function
$f_1(x^m)=\min_{j=1}^m (x_j)$. Assume that the network has a star
topology with node $1$ as the central node as illustrated in
Figure~\ref{fig:starnetwork}. Specifically, let $\mathcal{E} =
\{(j,1),(1,j)\}_{j=2}^m$.

Consider non-interactive coding schemes in which information is
constrained to flow in only one direction from the leaf nodes to the
central node as illustrated in Figure~\ref{fig:starnetwork}(a).
Specifically, the interaction protocol is given by $\mathcal{E}_i =
\{(j,1)\}_{j=2}^{m}$ for each $i =1, \ldots, t$. Since information
flows in only one direction from the leaf nodes to the central node,
there is no loss of generality in assuming that $t = 1$.  For each
$j=2,\ldots,m$, let us compute the cut-set bound $R_{sum,t}^{\mathcal
  S, \mathcal S^c}$ with $\mathcal{S}=\{j\}$ and $t = 1$.  Using
Lemma~\ref{lem:Han}, we obtain $R_{j1} \geq
H(X_j|X_1,\ldots,X_{j-1},X_{j+1},\ldots,X_m)=H(X_j)=1$. Therefore,
$R_{sum,1}\geq (m-1)$. Since this is achievable by transferring all
the data to node $1$, $R_{sum,1} = (m-1)=\Theta(m)$. Thus, in this
example, $t^* = 1$.

Now consider the following $(2m-2)$-round interactive coding scheme in
which information flows in both directions from the leaf nodes to the
central node and back as illustrated in
Figure~\ref{fig:starnetwork}(b). In round number $(2i-1)$, where $i$
ranges through integers from $1$ through $(m-1)$, node~$1$ sends the
sequence $\left(\min_{j=1}^i (X_j(k))\right)_{k=1}^n$ to node~$(i+1)$
at the rate $R_{1(i+1)(2i-1)}=H(\min_{j=1}^i (X_j))=h_2(1/2^i)$ bits
per sample. Node $(i+1)$ then computes the sequence
$\left(\min_{j=1}^{i+1} (X_j(k))\right)_{k=1}^n$ and sends it back to
node~$1$ in round number $2i$, using Slepian-Wolf coding (or
conditional coding) with the previous message as correlated side
information available to the decoder (and the encoder). This can be
done at the rate given by $R_{(i+1)1(2i)}=H(\min_{j=1}^{i+1}
(X_j)|\min_{j=1}^{i} (X_j))=1/2^i$ bits per sample. It can be verified
that the message sequence in round number $(2m-2)$ is the desired
function.

The sum-total-rate of this scheme is given by
\[\sum_{i=1}^{m-1}\left(h_2\left(
\frac{1}{2^i}\right)+\frac{1}{2^i}\right)\leq
\sum_{i=1}^{m-1}\left(\frac{1}{2^i}\log_2
e+\frac{i}{2^i}+\frac{1}{2^i} \right) <3+\log_2 e,
\]
where the first inequality is because $h_2(p)\leq p \log_2(e/p)$.
Thus for all $m\geq 6$, $R_{sum,(2m-2)} < 3+\log_2 e <5 \leq
(m-1)=R_{sum,t^*}$, showing that interaction is useful. In fact, the
minimum sum-total-rate $R_{sum,(2m-2)}$ is $O(1)$ with respect to the
number of nodes $m$ in the network. This is orderwise smaller than
$\Theta(m)$ for any $1$-round non-interactive coding
scheme.\footnote{Note the following: a) In studying how the minimum
sum-total-rate scales with network size, the coding blocklength is out
of the picture because it has already been ``sent to infinity''.  b)
Even though $H(\min_{j=1}^m (X_j)) \rightarrow 0$ as $m \rightarrow
\infty$, we cannot have nodes send nothing ($t = 0$) and set the
output of node $1$ to be identically zero. This is because then the
probability of block error will be equal to one.}

The above examples can be interpreted in two ways. From the
perspective of protocol design, these examples show that for a given
topology, certain information-routing configurations are fundamentally
more efficient than certain others for function computation. From the
perspective of network architecture, these examples show that certain
topologies are fundamentally more efficient than certain others for
function computation. The last example shows that the scaling laws
governing the information transport efficiency in large networks can
be dramatically different depending on whether the information
transport is interactive or non-interactive.

\section{Concluding remarks}

In this paper, we studied the two-terminal interactive function
computation problem within a distributed source coding framework and
demonstrated that the benefit of interaction depends on both the
function-structure and the distribution-structure.  We formulated a
multiterminal interactive function computation problem and
demonstrated that interaction can change the scaling law of
communication efficiency in large networks. There are several
directions for future work. In two-terminal interactive function
computation, a computable characterization of the infinite-message
minimum sum-rate is still open. The achievable infinite-message
sum-rate of Section~\ref{subsec:integralrate} involving definite
integrals and a rate-allocation curve appears to be a promising
approach. We have obtained only a partial characterization of the
structure of functions and distributions for which interaction is not
beneficial. An interesting direction would be to find necessary and
sufficient conditions under which interaction is useful. The
multiterminal interactive function computation problem is wide open. A
promising direction would be to study how the total network rate
scales with network size and understand how it is related to the
network topology, the function structure, and the
distribution-structure.


\appendices
\renewcommand{\theequation}{\thesection.\arabic{equation}}
\setcounter{equation}{0}

\section{\label{app:converse}Theorem~\ref{thm:rateregion} converse proof}

If a rate tuple $\textbf {R}= (R_1, \ldots, R_t)$ is admissible for
the $t$-message interactive function computation with initial
location $A$, then $\forall \epsilon > 0$, there exists
$N(\epsilon,t)$, such that $\forall n> N(\epsilon,t)$ there exists
an interactive distributed source code with initial location $A$ and
parameters $(t,n, |\mathcal{M}_1|, \ldots, |\mathcal{M}_t|)$
satisfying
\begin{eqnarray*}
&\frac{1}{n}\log_2 |\mathcal{M}_j|  \leq  R_j+\epsilon,\ \  j=1,\ldots, t, &\\
&\pP(\mathbf Z_A \neq \mathbf{\widehat Z}_A)  \leq  \epsilon, \
\pP(\mathbf Z_B \neq \mathbf{\widehat Z}_B) \leq  \epsilon.&
\end{eqnarray*}
Define auxiliary random variables $\forall i=1,\ldots,n, \ U_1(i)
:=\{M_1,X(i-),Y(i+)\}$, and for $j=2,\ldots,t$, $U_j := M_j$. \\

\noindent{\em Information inequalities:}
\noindent For the first rate, we have
\begin{eqnarray}
\lefteqn{n(R_1+\epsilon)}\nonumber \\
 &\geq& H(M_1) \nonumber\\
& \geq & H(M_1 | \mathbf{Y}) \nonumber\\
&\geq & I(M_1; \mathbf {X} | \mathbf{Y}) \nonumber\\
&=&H(\mathbf{X}|\mathbf{Y})-H(\mathbf{X}|M_1, \mathbf{Y}) \nonumber\\
&=&\sum_{i=1}^n H(X(i)|Y(i)) - H(X(i)|X(i-), M_1, \mathbf{Y}) \nonumber\\
&\geq & \sum_{i=1}^n H(X(i)|Y(i)) - H(X(i)|X(i-), M_1, Y(i), Y(i+)) \nonumber\\
&=&\sum_{i=1}^n I(X(i); M_1,X(i-),Y(i+)|Y(i)) \nonumber\\
&=&\sum_{i=1}^n I(X(i);U_1(i)|Y(i)). \label{eqn:rate1}
\end{eqnarray}

For an odd $j \geq 2$, we have
\begin{eqnarray}
\lefteqn{n(R_j+\epsilon)}\nonumber \\
 &\geq& H(M_j) \nonumber\\
&\geq & H(M_j | M^{j-1},\mathbf{Y}) \nonumber\\
&\geq & I(M_j; \mathbf {X} | M^{j-1},\mathbf{Y}) \nonumber\\
&=&H(\mathbf{X}|M^{j-1}, \mathbf{Y})-H(\mathbf{X}|M^j, \mathbf{Y}) \nonumber\\
&=&\sum_{i=1}^n H(X(i)|X(i-),M^{j-1},\mathbf{Y})-H(X(i)|X(i-),M^{j},\mathbf{Y}) \nonumber\\
&\stackrel{(a)}{=}&\sum_{i=1}^n H(X(i)|X(i-),M^{j-1},Y(i),Y(i+)) \nonumber\\&&-H(X(i)|X(i-),M^{j},\mathbf{Y}) \nonumber\\
&\geq &\sum_{i=1}^n H(X(i)|X(i-),M^{j-1},Y(i),Y(i+)) \nonumber\\&&-H(X(i)|X(i-),M^{j},Y(i),Y(i+)) \nonumber\\
&=&\sum_{i=1}^n I(X(i);M_j|M^{j-1},X(i-),Y(i+), Y(i)) \nonumber\\
&=&\sum_{i=1}^n I(X(i);U_j|U_1(i),U_2^{j-1}, Y(i)).
\label{eqn:rateodd}
\end{eqnarray}
Step (a) is because the Markov chain $X(i)- (M^{j-1}, X(i-), Y(i),
Y(i+))- Y(i-)$ holds for each $i=1,\ldots,n$.

Similarly, for an even $j \geq 2$, we have
\begin{eqnarray}
\lefteqn{n(R_j+\epsilon)}\nonumber\\
 &\geq& H(M_j) \nonumber\\
&\geq & I(M_j; \mathbf {Y} | M^{j-1},\mathbf{X}) \nonumber\\
&=&H(\mathbf{Y}|M^{j-1}, \mathbf{X})-H(\mathbf{Y}|M^j, \mathbf{X}) \nonumber\\
&=&\sum_{i=1}^n H(Y(i)|Y(i+),M^{j-1},\mathbf{X})-H(Y(i)|Y(i+),M^{j},\mathbf{X}) \nonumber\\
&\stackrel{(b)}{=}&\sum_{i=1}^n H(Y(i)|Y(i+),M^{j-1},X(i),X(i-))
\nonumber\\
&&-H(Y(i)|Y(i+),M^{j},\mathbf{X}) \nonumber\\
&\geq &\sum_{i=1}^n H(Y(i)|Y(i+),M^{j-1},X(i),X(i-)) \nonumber\\
&&-H(Y(i)|Y(i+),M^{j},X(i),X(i-)) \nonumber\\
&=&\sum_{i=1}^n I(Y(i);M_j|M^{j-1},X(i-),Y(i+), X(i)) \nonumber\\
&=&\sum_{i=1}^n I(Y(i);U_j|U_1(i),U_2^{j-1},
X(i)).\label{eqn:rateeven}
\end{eqnarray}
Step (b) is because the Markov chain $Y(i)-(M^{j-1}, X(i-), X(i),
Y(i+))- X(i+)$ holds for each $i=1,\ldots,n$.

By the condition $\pP(\mathbf Z_A \neq \mathbf{\widehat Z}_A) \leq
\epsilon$ and Fano's inequality\cite{CoverThomas},
\begin{eqnarray}
\lefteqn{h_2(\epsilon)+ \epsilon \log_2 (|\mathcal {Z}_A^n|-1)} \nonumber\\
 &\geq& H(\mathbf{Z}_A| M^t, \mathbf{X})  \nonumber\\
&\geq & \sum_{i=1}^n H(Z_A(i)| Z_A(i+), M^t, \mathbf{X})  \nonumber\\
&\geq & \sum_{i=1}^n H(Z_A(i)| Z_A(i+), Y(i+), M^t, \mathbf{X})  \nonumber\\
&\stackrel{(c)}{=}& \sum_{i=1}^n H(Z_A(i)| Y(i+), M^t, \mathbf{X})  \nonumber\\
&\stackrel{(d)}{=}& \sum_{i=1}^n H(Z_A(i)| Y(i+), M^t, X(i-),X(i))  \nonumber\\
&=&\sum_{i=1}^n H(Z_A(i)|U_1(i),U_2^t, X(i)). \label{eqn:ZA}
\end{eqnarray}
Step (c) is because for each $i$, $Z_A(i)=f_A(X(i),Y(i))$. Step (d)
is because the Markov chain $Z_A(i) - (X(i),Y(i)) - (M^t,
X(i-),X(i),Y(i+)) - X(i+)$ holds for each $i$. Similarly we also
have
\begin{equation}
h_2(\epsilon)+ \epsilon \log_2 (|\mathcal {Z}_B^n|-1)\geq
\sum_{i=1}^n H(Z_B(i)|U_1(i),U_2^t, Y(i)).\label{eqn:ZB}
\end{equation}

\noindent{\em Timesharing:} Then we introduce a timesharing random
variable $Q$ taking value in $\{1,\ldots,n\}$ equally likely, which is
independent of all the other random variables. Defining
$U_1:=(U_1(Q),Q), X:=X(Q), Y:=Y(Q), Z_A:=Z_A(Q), Z_B:=Z_B(Q)$, we can
continue (\ref{eqn:rate1}) as
\begin{eqnarray}
R_1+\epsilon & \geq & \frac{1}{n} \sum_{i=1}^n I(X(i);U_1(i)|Y(i)) \nonumber\\
&=&I(X(Q);U_1(Q)|Y(Q),Q)\nonumber\\ &\stackrel{(e)}{=}& I(X(Q);U_1(Q),Q|Y(Q))\nonumber\\
&=& I(X;U_1|Y), \label{eqn:rate1timesharing}
\end{eqnarray}
where step (e) is because $Q$ is independent of all the other random
variables and the joint pmf of $(X(Q),Y(Q))\sim p_{XY}$ does not
depend on $Q$.  Similarly, (\ref{eqn:rateodd}) and
(\ref{eqn:rateeven}) become
\begin{equation}
R_j+\epsilon \geq \left\{
\begin{array}{cc}
I(X;U_j|Y, U^{j-1}), & j \geq 2, j\mbox{ odd}, \\
I(Y;U_j|X, U^{j-1}), & j \geq 2, j\mbox{ even}.
\end{array}
\right.\label{eqn:ratejtimesharing}
\end{equation}
(\ref{eqn:ZA}) and (\ref{eqn:ZB}) become
\begin{eqnarray}
\frac{1}{n}h_2(\epsilon)+ \epsilon \log_2 |\mathcal {Z}_A| &\geq&
H(Z_A|U^t, X),\label{eqn:ZAtimesharing} \\
\frac{1}{n}h_2(\epsilon)+ \epsilon \log_2 |\mathcal {Z}_B| &\geq&
H(Z_B|U^t, Y).\label{eqn:ZBtimesharing}
\end{eqnarray}

Concerning the Markov chains, we can verify that $U_1(i)-X(i)-Y(i)$
holds for each $i=1,\ldots,n$, $\Rightarrow I(U_1(Q);Y(Q)|X(Q),Q)=0
\Rightarrow I(U_1(Q),Q;Y(Q)|X(Q))=0 \Rightarrow I(U_1;Y|X)=0$. For
each odd $j\geq 2$, we can verify that $U_j -
(X(i),U_1(i),U_2^{j-1})-Y(i)$ holds for each $i$, $\Rightarrow
I(U_j;Y(Q)|X(Q),U_1(Q),U_2^{j-1}, Q)=0 \Rightarrow
I(U_j;Y|X,U^{j-1})=0$. Similarly, we can prove the Markov chains for
even $j$'s. So we have
\begin{eqnarray}
\begin{array}{cc}
I(U_j;Y|X,U^{j-1})=0, & j\mbox{ odd}, \\
I(U_j;X|Y,U^{j-1})=0, & j\mbox{ even}.
\end{array}
\label{eqn:markovchains}
\end{eqnarray}

\noindent{\em Cardinality bounds:} The cardinalities $|\mathcal{U}_j|,
j = 1, \ldots, t$, can be bounded as in (\ref{eqn:cardinality}) by
counting the constraints that the $U_j$'s need to satisfy and applying
the Carath\'eodory theorem recursively as explained below (also see
\cite{Vis}). Let $U^t$ be a given set of random variables satisfying
(\ref{eqn:rate1timesharing}) to (\ref{eqn:markovchains}). If
$|\mathcal{U}_j|$, $j = 1, \ldots, t$ are larger than the alphabet
sizes given by (\ref{eqn:cardinality}), it is possible to derive an
alternative set of random variables satisfying (\ref{eqn:cardinality})
while preserving the values on the right side of
(\ref{eqn:rate1timesharing}) to (\ref{eqn:ZBtimesharing}) fixed by the
given $U^t$ as well as all the Markov chains (\ref{eqn:markovchains})
satisfied by the given $U^t$.  The derivation of an alternative set of
random variables from $U^t$ has a recursive structure. Suppose that
for $j = 1, \ldots, (k-1)$, alternative $U_j$ have been derived
satisfying (\ref{eqn:cardinality}) without changing the right sides of
(\ref{eqn:rate1timesharing}) to (\ref{eqn:ZBtimesharing}) and without
violating the Markov chain constraints (\ref{eqn:markovchains}). We
focus on deriving an alternative random variable $\widetilde U_k$ from
$U_k$. We illustrate the derivation for only an odd-valued $k$.  The
joint pmf of $(X,Y,U^t)$ can be factorized as
\begin{equation}
p_{XYU^t} = p_{U_k} p_{X U^{k-1}|U_k} p_{Y|X U^{k-1}} p_{U_{k+1}^t | X
Y U^k} \label{eqn:cardfactorization}
\end{equation}
due to the Markov chain $U_k - (X,U^{k-1})-Y$.  It should be noted
that $Z_A$ and $Z_B$ being deterministic functions of $(X,Y)$ are
conditionally independent of $U^t$ given $(X,Y)$.  The main idea is to
alter $p_{U_k}$ to $p_{\widetilde U_k}$ keeping fixed all the other
factors on the right side of (\ref{eqn:cardfactorization}). We alter
$p_{U_k}$ to $p_{\widetilde U_k}$ in manner which leaves
$p_{XYU^{k-1}}$ unchanged while simultaneously preserving the right
sides of (\ref{eqn:rate1timesharing}) to (\ref{eqn:ZBtimesharing}).
Leaving $p_{XYU^{k-1}}$ unchanged ensures that the Markov chain
constraints (\ref{eqn:markovchains}) continue to hold for
$U^{k-1}$. Fixing all the factors in (\ref{eqn:cardfactorization})
except the first ensures that the Markov chain constraints
(\ref{eqn:markovchains}) continue to hold for
$(\widetilde{U}_k,U_{k+1}^{t})$. To keep $p_{XYU^{k-1}}$ unchanged, it
is sufficient to keep $p_{XU^{k-1}}$ unchanged because
$p_{Y|XU^{k-1}}$ is kept fixed in
(\ref{eqn:cardfactorization}). Keeping $p_{XU^{k-1}}$ and $p_{X
U^{k-1}|U_k}$ fixed while altering $p_{U_k}$ requires that
\begin{equation}
p_{X U^{k-1}}(x,u^{k-1}) = \sum_{u_k} p_{U_k}(u_k) p_{X
U^{k-1}|U_k}(x,u^{k-1}|u_k)\label{eqn:cardmarginal}
\end{equation}
hold all tuples $(x,u^{k-1})$. This leads to $\left(|\mathcal{X}|
\prod_{j=1}^{k-1} |\mathcal{U}_j|-1 \right)$ linear constraints on
$p_{U_k}$ (the minus one is because $\sum_{x,u^{k-1}}p_{X
U^{k-1}}(x,u^{k-1}) = 1$). With $p_{XYU^{k-1}}$ unchanged, the right
sides of (\ref{eqn:rate1timesharing}) and (\ref{eqn:ratejtimesharing})
for $j =1,\ldots, (k-1)$ also remain unchanged. For $j=k$, $k$ odd,
the right side of (\ref{eqn:ratejtimesharing}) can be written as
follows
\begin{eqnarray}
i_{k} = H(X|Y,U^{k-1})- \sum_{u_k} p_{U_k}(u_k) H(X|Y,U^{k-1},U_k=u_k).
\label{eqn:c1}
\end{eqnarray}
The quantity $i_k$ is equal to the value of $I(X;U_k|Y,U^{k-1})$
evaluated for the original set of random variables $U^t$ which did not
satisfy the cardinality bounds (\ref{eqn:cardinality}). The quantities
$H(X|Y,U^{k-1})$, and $H(X|Y,U^{k-1},U_k=u_k)$ in (\ref{eqn:c1}) are
held fixed because $p_{XYU^{k-1}}$ is kept unchanged and all factors
except the first in (\ref{eqn:cardfactorization}) are fixed. In a
similar manner, for each $j > k$, $j$ odd, the right side of
(\ref{eqn:ratejtimesharing}) can be written as follows
\begin{eqnarray}
i_j = \sum_{u_k} p_{U_k}(u_k) I(X;U_j|Y,U^{k-1},U_{k+1}^j, U_k=u_k),
\label{eqn:c2}
\end{eqnarray}
where $i_j$ is equal to the value of $I(X;U_j|Y,U^{j-1})$ evaluated
for the original $U^t$ and $I(X;U_j|Y,U^{k-1},U_{k+1}^j, U_k=u_k)$ is
held fixed for all $j > k$, $j$ odd, because all factors except the
first in (\ref{eqn:cardfactorization}) are fixed. Again, for each $j >
k$, $j$ even, the right side of (\ref{eqn:ratejtimesharing}) can be
written as follows
\begin{eqnarray}
i_j = \sum_{u_k} p_{U_k}(u_k) I(Y;U_j|X,U^{k-1},U_{k+1}^j, U_k=u_k),
\label{eqn:c3}
\end{eqnarray}
where $i_j$ is equal to the value of $I(Y;U_j|X,U^{j-1})$ evaluated
for the original $U^t$ and $I(Y;U_j|X,U^{k-1},U_{k+1}^j, U_k=u_k)$ is
held fixed for all $j > k$, $j$ even, because all factors except the
first in (\ref{eqn:cardfactorization}) are fixed. The right sides of
(\ref{eqn:ZAtimesharing}) and (\ref{eqn:ZBtimesharing}) respectively
can also be written as follows
\begin{eqnarray}
h_A = \sum_{u_k} p_{U_k}(u_k) H(Z_A|X,U^{k-1},U_{k+1}^t,
U_k=u_k), \label{eqn:c4} \\
h_B = \sum_{u_k} p_{U_k}(u_k) H(Z_A|X,U^{k-1},U_{k+1}^t,
U_k=u_k), \label{eqn:c5}
\end{eqnarray}
where $h_A$ and $h_B$ are respectively equal to the values of
$H(Z_A|X,U^t)$ and $H(Z_B|Y,U^t)$ evaluated for the original $U^t$ and
$H(Z_A|X,U^{k-1},U_{k+1}^t, U_k=u_k)$ and $H(Z_A|X,U^{k-1},U_{k+1}^t,
U_k=u_k)$ are held fixed because $Z_A$ and $Z_B$ are deterministic
functions of $(X,Y)$ and all factors except the first in
(\ref{eqn:cardfactorization}) are fixed.

Equations (\ref{eqn:c1}) through (\ref{eqn:c5}) impose $(t-k+3)$
linear constraints on $p_{U_k}$. When the linear constraints imposed
by (\ref{eqn:cardmarginal}) are accounted for, altogether there are
no more than $\left(|\mathcal{X}| \prod_{j=1}^{k-1}
|\mathcal{U}_j|+t -k +2 \right)$ linear constraints on $p_{U_k}$.
The vector $(\{p_{XU^{k-1}}(x,u^{k-1})\},i_k,\ldots,i_t, h_A, h_B )$
belongs to the convex hull of $|\mathcal{U}_k|$ vectors whose
$\left(|\mathcal{X}| \prod_{j=1}^{k-1} |\mathcal{U}_j|+t -k +2
\right)$ components are given by
$\{p_{XU^{k-1}|U_k}(x,u^{k-1}|u_k)\}$, $H(X|Y,U^{k-1},U_k=u_k)$,
$\{I(X;U_j|Y,U^{k-1}, U_{k+1}^j, U_k=u_k)\}_{j > k, j:odd}$,
$\{I(Y;U_j|X,U^{k-1}, U_{k+1}^j, U_k=u_k)\}_{j > k, j:even}$,
$H(Z_A|X,U^{k-1},U_{k+1}^t, U_k=u_k)$,
$H(Z_A|X,U^{k-1},U_{k+1}^t,U_k=u_k)$. By the Carath\'eodory theorem,
$p_{U_k}$ can be replaced by $p_{\widetilde U_k}$ such that the new
random variable $\widetilde {U}_k \in \widetilde{\mathcal U}_k$
where $\widetilde{\mathcal U}_k \subseteq {\mathcal U}_k$ contains
only $\left(|\mathcal{X}| \prod_{j=1}^{k-1} |\mathcal{U}_j|+t-k+3
\right)$ elements, while (\ref{eqn:markovchains}) and the right
sides of (\ref{eqn:rate1timesharing}) to (\ref{eqn:ZBtimesharing})
remain unchanged.

\noindent{\em Taking limits:} Thus far, we have shown that $\forall
\epsilon >0$ and $\forall n> N(\epsilon,t)$, $\exists~
p_{U^t|XY}(u^t|x,y,\epsilon,n)$ such that $U^t$ satisfy
(\ref{eqn:cardinality}) and (\ref{eqn:rate1timesharing}) to
(\ref{eqn:markovchains}). It should be noted that
$p_{U^t|XY}(u^t|x,y,\epsilon,n)$ may depend on $(\epsilon,n)$, whereas
$\forall j=1,\ldots,t$, $|\mathcal U_j|$ is finite and independent of
$(\epsilon,n)$.  Therefore, for each $(\epsilon_0, n_0)$,
$p_{U^t|XY}(u^t|x,y,\epsilon_0,n_0)$ is a finite dimensional
stochastic matrix taking values in a compact set. Let $\{\epsilon_l\}$
be any sequence of real numbers such that $\epsilon_l>0$ and
$\epsilon_l \rightarrow 0$ as $l\rightarrow \infty$. Let $\{n_l\}$ be
any sequence of blocklengths such that $n_l > N(\epsilon_l,t)$. Since
$p_{U^t|XY}$ lives in a compact set, there exists a subsequence of
$\{p_{U^t|XY}(u^t|x,y,\epsilon_l,n_l)\}$ converging to a limit
$p_{\bar U^t|XY}(u^t|x,y)$. Denote the auxiliary random variables
derived from the limit pmf by $\bar U^t$. Due to the continuity of
conditional mutual information and conditional entropy measures,
(\ref{eqn:rate1timesharing}) to (\ref{eqn:markovchains}) become
\begin{equation*}
R_j \geq \left\{
\begin{array}{ccc}
I(X;\bar U_{j}|Y, \bar U^{j-1}), & I(\bar U_j;Y|X,\bar U^{j-1})=0, & j\mbox{ odd}, \\
I(Y;\bar U_{j}|X, \bar U^{j-1}), & I(\bar U_j;X|Y,\bar U^{j-1})=0, &
j\mbox{ even},
\end{array}
\right.
\end{equation*}
\[H(Z_A|\bar U^t, X)=0, \ \ H(Z_B|\bar U^t, Y)=0.\]
Therefore $\mathbf {R}$ belongs to right side of
(\ref{eqn:rateregion}). \hspace*{\fill}~\QED

\noindent {\em Remarks:} The convexity of the theoretical
characterization of the rate region can be established in a manner
similar to the timesharing argument in the above proof. The closedness
of the region can also be shown established in a manner similar to the
limit argument in the last paragraph of the above proof using the
following facts: (i) All the alphabets are finite, thus $p_{U^t|XY}$
takes values in a compact set. Therefore the limit point of a sequence
of conditional probabilities exists. (ii) Conditional mutual
information measures are continuous with respect to the probability
distributions.


\section{\label{app:lemmaonerecon}Proofs of
  Theorems~\ref{thm:onereconnohelp} and \ref{thm:nohelpgeneral}}

\noindent {\em Proof of Theorem~\ref{thm:onereconnohelp}:}
We need to show that $R_{sum,t}^A\geq H(X|Y)$ only for $p,q \in
(0,1)$. If $p,q \in (0,1)$ then $p_{XY}(x,y) > 0, \forall (x,y)\in
\mathcal X \times \mathcal Y$. Let $U^t$ be any set of auxiliary
random variables in (\ref{eqn:minsumrate}) satisfying all the Markov
chain and conditional entropy constraints of
(\ref{eqn:rateregion}). Due to Lemma~\ref{lem:rectangle}(i), for any
$u^t$, $\mathcal A(u^t)$ is a rectangle of $\mathcal X \times \mathcal
Y$. Due to Lemma~\ref{lem:rectangle}(iii) and the assumption that
$f_B(0,y_0)\neq f_B(1,y_0)$, $\mathcal A(u^t)$ cannot be $\mathcal X
\times \mathcal Y$. Therefore $\mathcal A(u^t)$ could be a row of
$\mathcal X \times \mathcal Y$, a column, a singleton, or the empty
set. Let
\[
\phi(u^t):=\left\{
\begin{array}{r@{,\  }l}
0 & \mbox{if } \mathcal A(u^t) \mbox{ is empty}\\
1 & \mbox{if } \mathcal A(u^t) \mbox{ is a row of } \mathcal X \times \mathcal Y \\
2 & \mbox{otherwise.} \end{array}\right.
\]

Now, $p_{U^t}(u^t)=0 \Leftrightarrow \mathcal A(u^t)$ is
empty. Therefore
\[
p_{\phi(U^t)}(0)= \sum_{u^t: \mathcal{A}(u^t) \mbox{ \footnotesize is
    empty}} p_{U^t}(u^t) = \sum_{u^t: p_{U^t}(u^t) = 0} p_{U^t}(u^t) =
0.
\]
Hence $p_{XY\phi(U^t)}(x,y,0) = 0$ for all $x$ and $y$. By the
definition of a row of $\mathcal X\times \mathcal Y$, we have
$H(X|U^t,\phi(U^t)=1)=0$, which implies that
$H(X|Y,U^t,\phi(U^t)=1)=0$. Similarly, we have
$H(Y|X,U^t,\phi(U^t)=2)=0$.  Loosely speaking, this means that knowing
the auxiliary random variables $U^t=u^t$ (representing the messages in
the proof of achievability), there are only two possible alternatives,
(1) $H(Y|X,U^t=u^t)=0$, that is, $\mathbf Y$ can be reproduced at
location $A$; (2) $H(X|Y,U^t=u^t)=0$, $\mathbf X$ can be reproduced at
location $B$.  Thus interestingly, although the goal was to only
compute a function of sources at location $B$, after $t$ messages have
been communicated, each location can, in fact, reproduce a part of the
source from the other location. In the case where $\mathbf X$ is not
known at location $B$, $\mathbf Y$ must be known at location $A$.

To continue the proof, for any $t \in \zZ^+$,
\begin{eqnarray*}
R^A_{sum,t} &=& \min [ I(X;U^t|Y)+I(Y;U^t|X)] \\
&=& \min[H(X|Y)-H(X|Y,U^t,\phi(U^t))\\&&+H(Y|X)-H(Y|X,U^t,\phi(U^t))]\\
&\stackrel{(a)}{=}& \min [H(X|Y)-H(X|Y,U^t,\phi(U^t)=2)p_{\phi(U^t)}(2)\\
&&+H(Y|X)-H(Y|X,U^t,\phi(U^t)=1)p_{\phi(U^t)}(1)\\
&=&\min [H(X|Y)-H(Y\oplus W|Y,U^t,\phi(U^t)=2)p_{\phi(U^t)}(2)\\
&&+H(Y|X)-H(X\oplus W|X,U^t,\phi(U^t)=1)p_{\phi(U^t)}(1)\\
&\geq &\min [H(X|Y)-H(W|\phi(U^t)=2)p_{\phi(U^t)}(2)\\
&&+H(Y|X)-H(W|\phi(U^t)=1)p_{\phi(U^t)}(1)\\
&=& \min [H(X|Y)+H(Y|X)-H(W|\phi(U^t))] \\
& \geq & H(X|Y)+H(Y|X)-H(W)\\
&\stackrel{(b)}{=}& H(X|Y),
\end{eqnarray*}
where all the minimizations above are subject to all the Markov
chain and conditional entropy constraints in (\ref{eqn:rateregion}).
In step $(a)$ we used the conditions $H(X|Y,U^t,\phi(U^t)=1)=0$ and
$H(Y|X,U^t,\phi(U^t)=2)=0$ and in step $(b)$ we used the fact that
$H(Y|X) = H(W) = h_2(p)$. \hspace*{\fill}~\QED

\noindent {\em Proof of Theorem~\ref{thm:nohelpgeneral}:} This follows
immediately by examining the proof of Theorem~\ref{thm:onereconnohelp}
and making the following observations. Observe that $\mathcal{A}(u^t)$
can be only a subset of a row or a column. This follows from the first
assumption in the statement of the theorem that these are the only
column-wise $f_B$-monochromatic rectangles of
$\mathcal{X}\times\mathcal{Y}$. Next observe that if
\[
\phi(u^t):=\left\{
\begin{array}{r@{,\  }l}
0 & \mbox{if } \mathcal A(u^t) \mbox{ is empty}\\
1 & \mbox{if } \mathcal A(u^t) \mbox{ is a subset of a row of } \mathcal X \times \mathcal Y \\
2 & \mbox{otherwise,} \end{array}\right.
\]
then $H(X|Y,U^t,\phi(U^t)=1)=0$ and $H(Y|X,U^t,\phi(U^t)=2)=0$ as in
the proof of Theorem~\ref{thm:onereconnohelp}. Finally observe that
the the series of information-inequalities in previous proof will
continue to hold if $X\oplus W$ and $Y\oplus W$ are replaced by
$\psi(X,W)$ and $\eta(Y,W)$ respectively. This is due to the second
assumption in the statement of the theorem which also states that
$H(Y|X) = H(W)$. \hspace*{\fill}~\QED

\section{\label{app:LBinfmsg}Theorem~\ref{thm:infinitemsgLB} proof}

Since $0<p,q<1$, $p_{XY}(x,y)>0, \forall (x,y)\in \mathcal X \times
\mathcal Y$. Let $U^t$ be any set of auxiliary random variables in
(\ref{eqn:minsumrate}) satisfying all the Markov chain and conditional
entropy constraints of (\ref{eqn:rateregion}). Due to
Lemma~\ref{lem:rectangle}(i) and (iv), for any $u^t$, $\mathcal
A(u^t)$ is a $f_A$-monochromatic rectangle of $\mathcal X \times
\mathcal Y$. Since $f_A(x,y)=x\wedge y$, $\mathcal A(u^t)$ can be
$\{(0,0),(0,1)\}$, $\{(0,0),(1,0)\}$, any singleton $\{(x,y)\}$, or
the empty set. Let
\[
\phi(u^t):=\left\{
\begin{array}{r@{,\ }l}
0 & \mbox{if } \mathcal A(u^t) \mbox{ is empty}\\
1 & \mbox{if } \mathcal A(u^t)=\{(1,1)\}\\
2 & \mbox{if } \mathcal A(u^t)\ni (1,0)\\
3 & \mbox{otherwise.} \end{array}\right.
\]

Since, $p_{U^t}(u^t)=0 \Leftrightarrow \mathcal A(u^t)$ is empty,
$p_{\phi(U^t)}(0)=0$. Therefore $p_{XY\phi(U^t)}(x,y,0) = 0$ for all
$x$ and $y$. When $X = Y = 0$, $\phi(U^t)$ can be only $2$ or $3$,
that is, $p_{XY\phi(U^t)}(0,0,0) = p_{XY\phi(U^t)}(0,0,1) = 0$. The
condition $p_{XY\phi(U^t)}(0,0,0) = 0$ is obvious. To see why
$p_{XY\phi(U^t)}(0,0,1) = 0$ is true, note that $\phi(u^t) = 1$ if,
and only if, $\mathcal{A}(u^t) = \{(1,1)\}$, which implies that
$p_{XYU^t}(0,0,u^t) = 0$ because $\mathcal{A}(u^t)$ is the set of all
$(x,y)$ for which $p_{XYU^t}(x,y,u^t) > 0$ and $(0,0)$ is not in it.
Therefore,
\[
p_{XY\phi(U^t)}(0,0,1) = \sum_{u^t: \mathcal{A}(u^t) = \{(1,1)\}}p_{XYU^t}(0,0,u^t)
= 0.
\]
Reasoning in a similar fashion, we can summarize the relationship
between $X,Y$, and $\phi(U^t)$ as shown in Table~\ref{tab:A0123}. For
each value of $(x,y)$, the values of $\phi(U^t)$ shown in the table
are those values for which $p_{XY\phi(U^t)}$ is possibly nonzero,
that is, for all values of $\phi$ different from those shown in the
table, the value of $p_{XY\phi(U^t)}$ is zero. For example, the
entry ``$X=0, Y=0, \phi(U^t)=2$ or $3$'' means that for $i \neq 2,3$,
$p_{XY\phi(U^t)}(0,0,i) = 0$.
\begin{table}[!htp]
  \centering
  \caption{Relation between $X,Y$ and $\phi(U^t)$}
  \begin{tabular}{|c|c|c|}
  \hline
   & $Y=0$ & $Y=1$ \\
   \hline
  $X=0$ & $\phi(U^t)=2$ or $3$ & $\phi(U^t)=3$ \\
  \hline
  $X=1$ & $\phi(U^t)=2$ & $\phi(U^t)=1$ \\
  \hline
\end{tabular}
\label{tab:A0123}
\end{table}

Let $\lambda:= p_{\phi(U^t)|X,Y}(2|0,0)$, we have
\begin{eqnarray*}
p_{\phi(U^t)}(0)&=&0,\\
p_{\phi(U^t)}(1)&=&p_{XY}(1,1),\\
p_{\phi(U^t)}(2)&=&p_{XY}(1,0)+\lambda p_{XY}(0,0), \\
p_{\phi(U^t)}(3)&=&p_{XY}(0,1)+(1-\lambda) p_{XY}(0,0).
\end{eqnarray*}
For any $t \in \zZ^+$,
\begin{eqnarray*}
\lefteqn{R^A_{sum,t}}\\ &=& \min [I(X;U^t|Y)+I(Y;U^t|X)] \\
&=& \min [I(X;U^t,\phi(U^t)|Y)+I(Y;U^t,\phi(U^t)|X)] \\
&\geq& \min [I(X;\phi(U^t)|Y)+I(Y;\phi(U^t)|X)]\\
&=& \min [H(X|Y)+H(Y|X)-H(X|Y,\phi(U^t))-H(Y|X,\phi(U^t))]\\
&=& \min \left[\right.h_2(p)+h_2(q)-H(X|Y=0,\phi(U^t)=2)p_{\phi(U^t)}(2) \\
&& -H(Y|X=0,\phi(U^t)=3)p_{\phi(U^t)}(3)\left.\right]\\
&\geq& \min_{0\leq \lambda\leq 1} \left[h_2(p)+h_2(q)-
h_2\left(\frac{p_{XY}(1,0)}{p_{\phi(U^t)}(2)}\right)p_{\phi(U^t)}(2) \right. \\
&&\left.-h_2\left(\frac{p_{XY}(0,1)}{p_{\phi(U^t)}(3)}\right)p_{\phi(U^t)}(3)\right],
\end{eqnarray*}
where all the minimizations above except the last one are subject to
all the Markov chain and conditional entropy constraints in
(\ref{eqn:rateregion}). The last expression is minimized when
$\lambda_* = q(1-p)/(p+q-2 p q)$. Evaluating the minimum value of the
objective function, we have
\[R_{sum,\infty} = \lim_{t\rightarrow \infty} R^A_{sum,t} \geq h_2(p)+h_2(q)-(1-pq) h_2\left(\frac{(1-p)(1-q)}{1-pq}\right).\]
\hspace*{\fill}~\QED

\footnotesize
\bibliography{mybibfile}


\end{document}